\documentstyle[12pt,a4,fleqn]{article}
\def\beq{\begin{equation}}
\def\eeq{\end{equation}}

\let\ka=\kappa
\let\la=\lambda
\let\La=\Lambda

\def\bl{\beta_{\la}}
\def\bk{\beta_{\ka}}
\def\bee2{\beta_{e^{2}}}

\def\pt{\partial_{t}}
\def\vl{\vec{\la}}
\def\vlN{\vec{\la}(N)}
\def\vl1{\vec{\la}_{1}}
\def\vl2{\vec{\la}_{2}}
\def\vb{\vec{\beta}}

\def\vb1{\vec{\beta}_{1}}
\def\vb2{\vec{\beta}_{2}}
\def\rt{\tilde{\rho}}
\def\gl#1{eq.~(\ref{#1})}
\def\s0#1#2{\mbox{\small{$\frac{#1}{#2}$}}}

\newsavebox{\tempbox}
\def\mycaption#1{
\sbox{\tempbox}{#1}
\vskip0.5cm
\ifdim \wd\tempbox >\hsize
#1
\else
\centering #1  
\fi
}

\newcommand{\be}{\begin{eqnarray}}
\newcommand{\ee}{\end{eqnarray}}

\newcommand{\G}{\mbox{$\Gamma$}}
\newcommand{\Gk}{\mbox{$\Gamma_k$}}

\renewcommand{\i}{\int\!}
\newcommand{\idx}{\int\!dx x^{\frac{d}{2}-1}}
\renewcommand{\d}{\delta }
\newcommand{\vp}{\mbox{$\varphi $ }}
\newcommand{\r}{\mbox{$\rho $ }}
\renewcommand{\l}{\mbox{$\la $ }}
\newcommand{\ls}{\mbox{$\la_\star $ }}
\newcommand{\es}{\mbox{$e^2_\star $ }}
\newcommand{\k}{\mbox{$\ka $ }}
\newcommand{\ks}{\mbox{$\ka_\star $ }}

\newcommand{\m}{\mu}
\newcommand{\n}{\nu}
\newcommand{\eb}{\mbox{$\bar{e}$}}

\newcommand{\Uk}{\mbox{$U_k$}}
\newcommand{\Ue}{\mbox{$U_k'$}}
\newcommand{\Uz}{\mbox{$U_k''$}}

\newcommand{\Zp}{\mbox{$Z_{\varphi,k}$}}
\newcommand{\Zf}{\mbox{$Z_{F,k}$}}
\renewcommand{\pd}[2]{\frac{\partial #1}{\partial #2}}
\newcommand{\grgl}{\:\hbox to -0.2pt{\lower2.5pt\hbox{$\sim$}\hss}
           {\raise3pt\hbox{$>$}}\:}
\newcommand{\klgl}{\:\hbox to -0.2pt{\lower2.5pt\hbox{$\sim$}\hss}
           {\raise3pt\hbox{$<$}}\:}

\begin{document}
\begin{flushright}
HD-THEP-94-49 \\
cond-mat/9502039
\\
\end{flushright}
\vspace{15mm}

\begin{center}
{\Large \bf{ Phase Transition of $N$-component Superconductors}}
\end{center}

\vspace{0.5cm}
\begin{center}
{\large B. Bergerhoff\footnote{e-mail:
bergerho@post.thphys.uni-heidelberg.de},
D. Litim\footnote{e-mail: cu9@ix.urz.uni-heidelberg.de},
S. Lola\footnote{e-mail: fe6@ix.urz.uni-heidelberg.de} and
C. Wetterich\footnote{e-mail: wetteric@post.thphys.uni-heidelberg.de}}
\end{center}
\vspace{-0.5cm}
\begin{center}
\begin{tabular}{c}
{\small Institut f\"{u}r
Theoretische Physik} \\ {\small Universit\"at
Heidelberg}\\
{\small Philosophenweg 16}\\{\small  69120 Heidelberg, Germany }\\
\end{tabular}
\end{center}

\vspace{1 cm}

\begin{center}
{\large {\bf Abstract}}
\end{center}

{\small We investigate
the phase transition in the
three-dimensional abelian Higgs model
for $N$ complex scalar fields, using
the gauge-invariant
average action $\Gamma_{k}$.
The dependence of $\Gamma_{k}$ on the
effective infra-red cut-off $k$ is described
by a non-perturbative flow equation. The
transition turns out to be
first- or second-order, depending on the ratio
between scalar and gauge coupling.
We look at the fixed points of the theory
for various $N$ and compute the
critical exponents of the model.
Comparison with results from the
$\epsilon$-expansion shows a
rather poor convergence for $\epsilon=1$
even for large $N$. This is in contrast to
the surprisingly good results of the $\epsilon$-expansion
for pure scalar theories. Our results suggest the existence
of a parameter range with a second-order transition
for all $N$, including the case of the
superconductor phase transition
for $N=1$.
}


\thispagestyle{empty}

\setcounter{page}{0}
\vfill\eject

\section{Introduction}

The quantitative details of the phase transition
in superconductors are so far not fully
understood.
If the phase transition is second-order or very weakly
first-order, it is believed that
the universal behaviour in the immediate vicinity
of the critical temperature can be described
by a field theory of electrodynamics
with a charged scalar field.
Here the scalar field may represent
composite degrees of freedom like
Cooper-pairs in ``standard'' superconductors.
A non-vanishing expectation value of
this scalar field indicates
spontaneous symmetry breaking and corresponds
to the superconducting phase.
Even though the critical behaviour for
low temperature superconductors may be difficult to
observe, this situation may improve for high temperature
superconductors
\cite{physicaA}.
Also, it is claimed that
certain phase transitions in liquid crystals
are described by the same universality class
as for the superconductors
\cite{HaLu}.
Critical exponents have been measured for these
transitions
\cite{Joetal}.
An understanding of the
critical behaviour in the
three-dimensional abelian Higgs model, which
should describe all these transitions,
is also interesting from the theoretical
side: This is the simplest example
with a continuous gauge symmetry.
Furthermore, the high
temperature phase transition in the four-dimensional
abelian Higgs model should be well
approximated in the vicinity of the phase transition
by a three-dimensional effective theory
\cite{DimRed}.
Here the abelian gauge theory may serve as
a prototype for the
understanding of the non-abelian model which describes the
electroweak phase transition in the early universe.

In this paper we address the problem of the phase transition
in the three-di\-men\-sio\-nal field theory of scalar electrodynamics.
Even the order of the
phase transition is not settled so far,
ranging from proposals of
a first-order transition
for all values of the coupling
constants
\cite{suco}
to speculations about a
possible second-order
behaviour
\cite{DAS}.
On the one side
high temperature perturbation theory
in the four-dimensional model seems to
predict a first-order
transition. Perturbation theory is,
however, plagued by severe infra-red divergences.
In the symmetric phase
the one-loop correction to
the quartic scalar coupling
diverges as $k^{-1}$, where $k$
is an appropriate infra-red cut-off.
Although
it is possible to improve
the calculations
by a resummation of diagrams,
there are questions on the reliability
of this method \cite{resum}.
Moreover, from the one-loop
correction to the gauge
coupling,
an infra-red divergence
$ \sim k^{-1}$ appears
if the scalar mass vanishes. Since this
is the case approximately near the phase transition,
one  concludes that
in this region the results of
perturbation theory will not be
accurate. The
encountered infra-red problems are exactly those
of three-dimensional perturbation theory.
In order to avoid this difficulty
an alternative approach uses the $\epsilon$-expansion
and attempts to extrapolate results of a computation
in $4-\epsilon$ dimensions to $\epsilon = 1$.
This has not given a convincing picture
so far.  The situation improves somewhat if,
in addition, one considers $N$ charged scalar
fields with $N$ large. The phase transition
is well understood for $N \rightarrow \infty$
but an extrapolation to $N=1$
is not easy.
We will see that even the leading
$1/N$ contribution
to the critical exponents is not
well described by the $\epsilon$-expansion.

In recent years,
a new method which deals with
these infra-red problems
in a non-perturbative way has been proposed
\cite{ch1}. The method is based
on the effective average action, $\Gk$, for
which only fluctuations with momenta
$q^{2} > k^{2}$ are included.
The scale $k$ acts as
an effective infra-red cut-off.
An exact, non-perturbative evolution
equation describes the scale
dependence of the average action
\cite{cwe}.
Using this equation, it is possible
to follow the $k$-dependence
of $\Gamma_{k}$ to extrapolate to the limit
$k \rightarrow 0$, where $\Gamma_{k}$ becomes the
generating functional for the
1PI Green functions.
The method has been tested in
the $\Phi^{4}$-theory \cite{ch2},
where it successfully
describes the second-order phase
transition that the system
undergoes. A study of the critical
behaviour of the three-dimensional
$\Phi^{4}$-theory gives values for
the critical exponents in very good
agreement
with previous estimates \cite{expo}. A gauge-invariant
effective average action
can be constructed for abelian and non-abelian
gauge theories
\cite{ch3}.
The corresponding non-perturbative
exact evolution equation
for the abelian Higgs model has been derived
for arbitrary dimension $d$
\cite{ch3}.
For $d = 4$ and small gauge coupling,
the flow equation is
similar to the perturbative
renormalization group equations
in the one-loop approximation.
A numerical study, based
on this work,
indicates
a first-order character
for the Coleman-Weinberg
phase transition
\cite{daniel}, in agreement
with the
perturbative analysis.
However, for $d=3$,
strong renormalisation effects
of the gauge coupling
result in predictions that
may not be obtained
by perturbation theory.
In particular, the
dimensionless coupling $e^{2}$
reaches in the scaling limit a value
of order $6 \pi^{2}$, thus
$e^{2}/16 \pi^{2}$ is not
a small quantity anymore
\footnote{
For the non-abelian
Higgs model in three dimensions
the dimensionless gauge coupling
also grows large. Additional
complications arise from confinement
effects \cite{bastian}.
}.

The order of the phase transition seems
to depend on the ratio $\lambda_0/e^2_0$
of the bare quartic scalar coupling $\lambda_0$
and the squared gauge coupling $e^2_0$.
For small $\lambda_0/e^2_0$
(type I superconductors) it was suggested that
the phase transition is first-order
\cite{suco}, the discontinuity being induced by
fluctuations of the gauge field. This picture has
been confirmed by the solution of the non-perturbative
flow equation \cite{ber}.
The question about what happens for large scalar couplings
has not found a clear answer by previous methods,
except for very large $N$ (typically $N \geq 200$)
where a second-order phase transition was established for
sufficiently large $\lambda_0/e^2_0$. Investigations based
on the average action indicate
the existence of a range in parameter space where the
transition is second-order also for the superconductor
$(N=1)$ \cite{ber}. If a parameter
range for a second-order transition
exists for all $N$ there must be a fixed point of the flow
equations with associated scale invariant critical behaviour.

In this paper we will
use the concept of the average action to
examine  the
critical behaviour
of the
three-dimensional abelian Higgs model
for $N$ complex scalar fields.
We will discuss how the fixed points of the theory
move as $N$ varies and
examine the
critical exponents of the model, in the
region where a second-order
phase transition arises.
In section 2 we give a brief
sketch of the basic formalism and derive the
evolution equations
for the running of the couplings.
In section 3 we look at the scaling solutions in
the large-$N$ approximation.
In section 4 we
discuss a differential equation which allows us
to follow the $N$-dependence of the
fixed points. This uses
a linearization
of the flow equation for values
of the couplings near the fixed point
values.
In section 5
we give a detailed numerical analysis
of the model for various $N$.
The
critical exponents for a second-order
phase transition are discussed in section 6.
In section 7 we present our results
and compare them with
those obtained through the $\epsilon$-expansion.
Finally in section 8 we give a
summary of the work and
derive the conclusions.

\section{Average action and
flow equations \newline
for running couplings}

The starting point
of the analysis
is the classical action $S$
for $N$ complex scalar fields $\chi_{a},
\; a = 1,..,N$
and the gauge field ${\cal{A_\m}}$
\be
S[\chi,{\cal{A_\m}}] = \i d^dx \left\{ \frac{1}{4}
{{\cal{F}}^{\m\n}{\cal{F}}_{\m\n}}\!+\!
\left(D^\m\chi_a\right)^*
\left(D_\m\chi^a\right)\!-\!
\bar{\m}^2\chi_a^* \chi^a\!+\!\frac{\bar{\la}}{2}\left(\chi_a^*
\chi^a\right)^2
\right\} ,
\label{3.2.1}
\ee
where $\cal{F}_{\m\n} = \partial_\m \cal{A}_\n -
\partial_\n \cal{A}_\m$,
$D_\m = \partial_\m + i \bar{e} \cal{A}_\m$.
We note that the pure scalar sector is
$O(2N)$-symmetric. The coupling to the photon
reduces this symmetry to
$SU(N) \times U(1)$ where the abelian
part is gauged.
Furthermore, we have the gauge fixing term
\be
S_{gf}
 = - \frac{1}{2\alpha} \i d^dx ({\cal A}^{\mu}-\bar{A}^{\mu})
\partial_\m
\partial_\n ({\cal A}^{\nu}-\bar{A}^{\nu}) ,
\label{3.3}
\ee
where  $\bar{A}_{\mu}$ is some arbitrary
background field.
The average action $\Gamma_k$ is the effective
action for (gauge-invariant) averages of
fields. It is obtained by integrating
out only quantum fluctuations
with (covariant) momenta larger than $k$.
For this purpose one introduces
in the functional integral for the usual effective
action an infra-red cut-off $R_{k}$. If $R_{k}$
vanishes for $k \rightarrow 0$
one recovers the generating functional for the
1PI Green functions $\Gamma$ as $ \lim_{k \rightarrow 0 }$
$\Gamma_{k}$.
The functional $\Gamma_{k} [\varphi, A]$ is gauge-invariant
for all $k$.
The dependence of the effective average action
on the scale $k$ is determined by an exact
non-perturbative flow equation
\cite{ch3}
which reads (up to minor modifications)
\be
\frac{\partial}{\partial t} \G_k[\vp\!,A]\!=\!
\frac{1}{2} \mbox{Tr}\left\{\!\left( \pd{R_k[A]}{t} \right)\!
\left( \G_k^{(2)}[\vp\!,A]
\!+\!\Gamma_{gf}^{(2)}
\!+\! R_k[A]\right)^{-1} \right\}
\!+\! \frac{\partial}{\partial t}C_k[A] .
\label{1.20}
\ee
Here the trace involves a momentum integration
and summation over internal indices,
$t = \ln k / \Lambda$,
and $\Gamma_{k}^{(2)}$ is the second
functional variation of
$\Gamma_{k}$. The sum $\Gamma_k^{(2)} +
\Gamma_{gf}^{(2)}$ corresponds to
to the
exact inverse propagator in presence of
the fields $\varphi$ and $A$.
We approximate $\Gamma_{gf}$ by
(\ref{3.3}) with
${\cal A}_{\mu}$ replaced by $A_{\mu}$.
More details can be found in \cite{ch3}
where also the correction $C_{k}[A]$ is
motivated.\\
The infra-red cut-off $R_k$
acts on scalars $(R_k^{(S)})$
and gauge fields $(R_k^{(G)})$
and reads explicitly
\be
R_k^{(S)} \!&=&\! \Zp \cdot \left(-D^2(A)\right)
\frac{f_k^2\left(-D^2(A)\right)}{1 - f_k^2\left(-D^2(A)\right)}
\nonumber\\
R_k^{(G)} \!&=&\! \Zf \cdot \left(-\partial^2\right)
\frac{f_k^2\left(-\partial^2\right)}{1 - f_k^2
\left(-\partial^2\right)} .
\label{3.4}
\ee
Here $\Zp$ and $\Zf$ are appropriate
wave function renormalization constants
for the scalar
and gauge field, respectively.
We choose a cut-off that decays exponentially
for large (covariant) momentum squared $x=q^2$, i.e.
\beq
f_k^2(x) = e^{-x/k^2} .
\eeq
For $x \rightarrow 0$ the cut-off
$R_k \rightarrow Z k^2$ becomes a mass-like term
and suppresses the propagation
of quantum fluctuations
with $q^2 < k^2$. This eliminates the
infra-red-divergence problem
that appears in the perturbative
approach. For large $q^2$
the exponential decay of
$\partial_t R_k$ guarantees the ultra-violet
convergence of the momentum
integral in (\ref{1.20}). Finally, the
use of the covariant derivative in (\ref{3.4})
ensures explicit gauge covariance of the flow
equation.

A discussion of the phase transition relies mainly on the
properties of the effective scalar potential
$U_{k}$ which corresponds to the free energy.
In order to extract from (\ref{1.20})
the flow equation for the average potential
$U_k$ we use a configuration with
$\vp = const.$
and $A=0$.
For the derivation of
the flow equation for $U_k$
we use a truncation of the
average action of the form
\beq
\Gamma_k[\varphi,A] =
\int d^d x \left (
\frac{1}{4}{Z_{F,k}}F_{\mu \nu}F^{\mu \nu}
+ Z_{\varphi,k} (D^{\mu}
\varphi_{a})^* D_{\mu}
\varphi^{a} + U_k(\rho) \right ) .
\label{trun}
\eeq
The second functional derivative of this expression
can be found in appendix A.
One obtains
for arbitrary dimension $d$ the flow equation
\be
\frac{\partial}{\partial t}\Uk (\rho)& = &(d-1) v_d \idx
\tilde{\partial}_t \ln \left(\Zf P(x) + 2 \Zp \eb^2 \r
\right)  \nonumber \\
&&+v_d \idx \tilde{\partial}_t \ln \left(\Zp P(x) + \Ue(\r) +
2 \r \Uz(\r) \right)  \nonumber\\
&&+(2N-1)v_d \idx \tilde{\partial}_t
\ln \left(\Zp P(x) + \Ue(\r) \right) ,
\label{pde}
\ee
where
\begin{eqnarray}
v_d^{-1} & = & {2^{d+1}\pi^{d/2}\Gamma(\frac{d}{2})} .
\end{eqnarray}
Here the potential is a function of the
invariant
$ \rho =  \varphi_a^* \varphi^a$
and primes denote derivatives with respect to
$\rho$.
We also use the variable
$x = q^2$ and the inverse average propagator
$P(x)$ is defined by
\begin{eqnarray}
Z P(x) & = & Z x + R_k(x)  \nonumber \\
P(x) & = & \frac{x}{1-\exp(-\frac{x}{k^2})} .
\end{eqnarray}
The partial derivative
$\tilde{\partial}_t$ acts only on the infra-red cut-off
$R_k$ in $P$ with
\beq
\tilde{\partial}_t P(x) =
k \frac{\partial}{\partial k} P(x) -
(k \frac{\partial}{\partial k} \ln Z_k)
(P(x)-x)
\label{df}
\eeq
and we drop the last term proportional to the
anomalous dimension in this paper.\\
In the sequel, the use of dimensionless
variables turns out to be convenient.
We therefore introduce the following variables:
\be
u_k(\rt)&=&k^{-d} U_k(\rho) \\
\rt &=&k^{2-d} \Zp \, \rho \\
e^2 &=& k^{d-4} (\Zf)^{-1} \bar{e}^2 .
\ee
In terms of the above, the partial
differential equation (\ref{pde}) can be rewritten as
\be
\pt u_k(\rt)&=& -d\, u_k(\rt) +
(d-2+\eta_\varphi)\rt\ u'_k(\rt) +
2(d-1)v_d l^d_0 s^d_0(2e^2\rt) \nonumber \\
&&  +2(2N-1)v_d l^d_0 s^d_0(u'_k(\rt)) +
2v_d l^d_0 s^d_0(u'_k(\rt)+2\rt u''_k(\rt)) \label{exact} ,
\ee
where primes denote now partial derivatives
with respect to $\rt$.  The anomalous dimension $\eta_\varphi = -
\partial_t \ln Z_{\varphi,k}$
reads
\be
\eta_{\varphi} = -16 v_d \left(1-\frac{1}{d}\right)e^2
l^d_{1,1}(2\la\ka,2 e^2\ka) +
\frac{16}{d} v_d \la^2\ka\ m^d_{2,2}(2\la\ka,0)\ ,
\label{andim}
\ee
and we introduced the dimensionless renormalized
vacuum expectation value \k defined through $u_k'(\ka)=0$
and the dimensionless renormalized quartic coupling \l as
\beq
 \la(k)=u''_k(\ka(k))\ .
\label{quarca}
\eeq
The constants $l^d_0$ and the threshold
functions $s^d_0(\omega)$,  $l^d_{n,m}(\omega_1,\omega_2)$ and
$m^d_{n,m}(\omega_1,\omega_2)$  are given in appendix B.
We also need an evolution equation for
$e^2(k)$ that we infer from \cite{ch3}.

The main subject of the present paper is
the investigation of solutions to \gl{exact}.
In order to solve \gl{exact} for $u_k(\rt)$
we make first an Ansatz for the effective potential $u_k(\rt)$  as
a polynomial expansion around the $k$-dependent
minimum at $\ka(k)$. The partial differential
equation (\ref{exact}) can then be transformed
into infinitely many coupled ordinary differential
equations for the higher derivatives of the
effective potential taken at the vacuum expectation
value. To lowest order we find the following coupled
set of differential equations for the regime
with spontaneous symmetry breaking
\be
\frac{de^2}{dt} \!  = &\! \bee2 =& (d-4)e^2 \nonumber\\
&&\! +\frac{4}{3}v_d e^4 \left[l^d_g
\tilde{s}^d_g(2 \la \ka,2e^2\ka) +
l^d_c(2 \la \ka) +(N-1)l^d_{gc}\right]
\label{betae2} \\
\frac{d \ka}{d t} \!=&\! \bk =
    &  \left( 2 - d - \eta_\varphi \right) \ka +
      4 \frac{e^2}{\la} (d-1) v_d
      l^d_1 s^d_1(2 e^2 \ka)  \nonumber \\
      & &\! +6
      v_d l^d_1 s^d_1(2 \la \ka)
      + 2 (2N-1) v_d l^d_1  \label{betaka} \\
\frac{d \la}{d t} \!=&\! \bl= &\left( d -
      4 + 2 \eta_\varphi \right) \la +
      8 e^4 \left( d-1 \right) v_d l^d_2
      s^d_2(2 e^2 \ka) \nonumber \\
      & &\!
      +18 \la^2
      v_d l^d_2 s^d_2(2 \la \ka) + 2
      (2N-1)\la^2 v_d l^d_2.
\label{betala}
\ee
The functions
$s^d_n(\omega)$,
$l_c^d(\omega)$ and
$\tilde{s}^d_{g}(\omega_1,\omega_2)$ describe the
threshold effect due to mass terms. They are given,
as well as the constants $l^d_n$, $l^d_{gc}$
and $l_g^d$, in appendix B.
We emphasize that our $\beta$-functions are directly
computed for arbitrary dimension $d$ and
can immediately be evaluated
for $d=3$.
No $\epsilon$-expansion around $d=4$ is necessary.
We will see later (sect.~7) that the
$\epsilon$-expansion for $\epsilon = 1$ fails to reproduce
our results.

The $N$-dependence in $\beta_{\ka}$
and $\beta_{\la}$ is coming from
diagrams that involve the $(2N-1)$ massless
scalar modes inside the loop.
We note that the scalar contributions are the same
as for a pure $SO(2N)$-symmetric scalar theory
for which one has $(2N-1)$ massless
Goldstone excitations around a minimum
at $\kappa \neq 0$.
For $\beta_{e^2}$ the $N$-dependence
arises from the diagrams that involve the
$(N-1)$ fields
$\tilde{\sigma}$ and $\tilde{\omega}$
(appendix A). They correspond to
$N-1$ complex massless scalars which have a vanishing
expectation value. These $2N-2$ real Goldstone modes
belong to the fundamental representations
of the unbroken global symmetry group
\mbox{$SU(N-1)$}. The contribution of the remaining
``would be'' Goldstone boson (singlet of \mbox{$SU(N-1)$})
is contained in the term for $N=1$.

Starting with sufficiently large initial values of
$\lambda / e^2$ and solving the system of
equations (\ref{betae2})-(\ref{betala})
for $k \rightarrow 0$ shows that $\kappa$ either
runs to zero at some non-vanishing scale
$k_s$ or diverges $\sim k^{-1}$. The first behaviour
corresponds to the symmetric phase where
$\rho_0(k=0) = 0$ whereas
the second indicates spontaneous symmetry breaking
with $\rho_0(k=0) = \rho_0 > 0$,
$\kappa = \rho_0 k^{-1}$. By an appropriate
tuning of the initial value of $\kappa$
the scales $k_s$ or $\rho_0$ can be made arbitrarily small. The
phase transition is second-order.
For the initial $\kappa$ taking exactly the critical value,
the solution of the flow equations runs asymptotically
towards a scaling solution
for $k \rightarrow 0$. This is described
by fixed point values $\kappa_\star$, $\lambda_\star$ and
$e^2_\star$ and corresponds to the scale invariant physics at
the phase transition. On the other side,
for a small ratio of initial values $\lambda / e^2$ the coupling
$\lambda$ may reach zero (or extremely small values)
for nonvanishing $k$. Looking at the flow equation
(\ref{exact}) for $u_{k}$ this corresponds to a situation where
already the absolute minimum of $u_k$ has jumped
to the origin at $\tilde{\rho} = 0$.
This behaviour indicates a first-order phase transition.

\section{Scaling solutions and large-$N$ approximation}

Thus, we find that the abelian Higgs model
in three dimensions has a region of parameter values
where the phase transition is first-order,
and a region where it is second-order.
This can be
qualitatively
understood by looking at the two limits of the theory:
When no gauge-field fluctuations are present
$(e^2=0)$ the model reduces to the $2N$-component
Heisenberg model with a second-order
phase transition. Below
the critical
temperature
the effective potential $U_0$ has a maximum at
$\rho = 0$ whereas the minimum occurs for $\rho_0=\rho_0(k=0)$.
On the other hand, gauge-field fluctuations
dominate the evolution equations
for $\la \ll e^2$. The solution
of the flow equation is then
similar to the four-dimensional model
\cite{daniel} and results in a first-order phase transition.
A local minimum remains at the origin even
somewhat below the critical temperature.

At the critical temperature
$T_c$ of a second-order phase transition
the theory becomes scale invariant. This corresponds
to a fixed point where the dimensionless parameters
$\ka$, $\la$ and $e^2$ take $k$-independent
values. In consequence, the unrenormalized
quantities $\rho_{0}$, $U''_k(\rho_0)$ and $\bar{e}^2$
vanish with appropriate powers of $k$.
This ``second-order fixed point'' is infra-red
stable in all couplings except
one ``relevant coupling'' that we may associate with $\ka$.
We specify short distance
(or microscopic) couplings at some appropriate
high momentum scale $\La$. Then the critical
surface, for which the phase transition
occurs, is given by a critical value
$\ka_{c}(\La)$ which is a function of
$\la(\La)$ and $e^2(\La)$.
The difference $\ka(\La)-
\ka_{c}(\La)$ is proportional to
$T_c-T$. With initial values on the critical
surface all couplings flow towards
the second-order fixed point if
one starts within its range of attraction.
This range of attraction also defines the region in
parameter space for which the transition is
second-order. There must also be a line
(more generally a hypersurface)
which separates this parameter region from
the one for which the transition becomes first-order.
On this line, the couplings flow
towards another fixed point,
the ``tricritical fixed point''.
This tricritical fixed point has two infra-red
unstable directions (in $\ka$ and $\la$).
Our picture of a parameter range
with second-order transition and a different range with
first-order transition is thus realized
if two non-trivial fixed points can be identified, one
with one and the other with two unstable directions.

We therefore want to investigate the scaling solutions related to
the fixed points of
the coupled system of differential equations
(\ref{betae2}) - (\ref{betala}) in three
dimensions, i.e. we look for simultaneous
zeros of the r.h.s. of eqs. (\ref{betae2}) - (\ref{betala}).
It turns out that the system (\ref{betae2}) - (\ref{betala}) has
non-trivial
fixed points for any $N$, although previous results
\cite{suco}, obtained through an $\epsilon$-expansion
method, seem to indicate that a large number
of complex fields would be needed
for this purpose.
The main new ingredient for the existence
of fixed point solutions even for small $N$
stems from the fact that threshold effects due
to the decoupling of massive fluctuations are properly
taken into account. As an illustration
\footnote{ We have chosen $N=10$, but
fig.~1 describes the generic
behaviour for any $N$ below $\sim 183$. For
$N > 183$, fixed points are present even
when the threshold effects are discarded. }
, we have
depicted in fig.~1
the $\beta$-function
for $\la$ (\ref{betala}) as
a function of $\la$ at fixed $e^2 = e^2_\star$.
Neglecting the threshold effects corresponds
formally to $\ka=0$ and results in the
upper curve. In this case, no fixed
point for $\la $ is obtained. Taking the
full threshold behaviour into account (with
$\ka$ at its fixed point $\ka_\star$)
reveals the occurrence of two fixed points for $\la$ (zeroes for
$\bl$). Here the smaller value of \l corresponds to the
tricritical and the larger one to
the second-order fixed point, respectively.

To start a quantitative discussion of the location of the
fixed points we consider in this section the large-$N$ limit.
The evolution equation for the gauge coupling decouples
now completely from the scalar sector and reads
\beq \label{be2n}
\frac{d e^2}{d t} \!=\! -e^2 +
       \frac{4}{3} N v_3 l^3_{gc} e^4 .
\eeq
Its non-trivial infra-red fixed point solution is given by
\beq \label{e2f}
e^2_\star=\frac{3}{4 v_3 l^3_{gc}}\frac{1}{N}.
\eeq
The corresponding solutions for the scalar
sector can be obtained through an Ansatz
for the large-$N$ behaviour of the couplings
as $\ka \sim N^{\alpha}, \la \sim N^{\beta}$.
Inserting this Ansatz in eqs.~(\ref{betaka})
and (\ref{betala}) it follows that only the pairs
\beq \label{BB1}
(\alpha,\beta)=(1,-1)\,;\,(1,-2) \nonumber
\eeq
may yield solutions to the fixed point equation.
This result obtains also through an appropriate
rescaling of the full partial differential
equation (\ref{exact}) and is therefore no mere effect
of the polynomial approximation.
Here $\la_\star \sim 1/N$ corresponds to the
second-order fixed point, whereas the tricritical point is
characterized by $\la_{\star} \sim 1/N^2$.
More explicitely we find for
the second-order fixed point
$(\beta = -1)$ that the influence of the gauge
coupling (\ref{e2f}) and of the massive scalar
fluctuation in the evolution equations
(\ref{betaka}) - (\ref{betala}) are of subleading
order in $1/N$. Therefore, they simplify to
\be
\frac{d \ka}{d t} \!&=&\! -\ka +
      4 N v_3 l^3_1
\label{b}
\ee
\be
\frac{d \la}{d t} \!&=&\!  -\la +
       4 N  v_3 l^3_2 \la^2,
\label{c}
\ee
with the corresponding  fixed point values given by
\be
\ka_\star \!&=&\! 4 v_3 l^3_1\,N \label{ka1f}
\ee
\be
\la_\star \!&=&\! \frac{1}{4 v_3 l^3_2}\, \frac{1}{N}.\label{la1f}
\ee
Since eqs.~(\ref{be2n}), (\ref{b}) and (\ref{c})
are decoupled, it is obvious that
the fixed point characterized by
eqs.~(\ref{e2f}),(\ref{ka1f}) and ({\ref{la1f})
is infra-red stable in $\la$ and $e^2$ and unstable in
$\ka$. Since it has only one unstable direction
this must be the second-order fixed point. The
fixed point values \ks and \ls
are entirely determined
through the
fluctuations of the massless modes and any memory
of the presence of the gauge field has disappeared
in this limit. Therefore, in the large-$N$ limit,
the fixed point
values $\la_\star$ and $\ka_\star$
are identical
to the Wilson-Fisher fixed point of the $O(2N)$-symmetric pure
scalar model.
This simple relation does not hold
true for the subleading terms
in $\la_\star$
nor for the critical indices of the theory.
The anomalous dimension, for example, reads
for the fixed point (\ref{e2f}),(\ref{ka1f}) and ({\ref{la1f}):
\beq
\eta_\star=\left[ \frac{4}{3} \frac{l^3_1}{(l^3_2)^2} \,
m^3_{2,2}(\tilde{m}^2_\star,0) -\frac{8}{l^3_{gc}}\,
l^3_{1,1}(\tilde{m}^2_\star,\tilde{M}^2_\star) \right] \frac{1}{N} ,
\label{etaf1}
\eeq
where $\tilde{m}_\star$ (resp.~$\tilde{M}_\star$)
denotes the dimensionless scalar- (gauge-) field mass
at the fixed point:
\be
\tilde{m}^2_\star&=& 2 \la_\star \ka_\star =  \frac{2l^3_1}{l^3_2}
= 1.7 \label{mf1} \\
\tilde{M}^2_\star&=& 2 e^2_\star \ka_\star =  \frac{6 l^3_1}{l^3_{gc}}
= 6.3 \,\, .
\label{Mf1} \ee
The first term in eq.~(\ref{etaf1}) stems
from the scalar fluctuations only (and corresponds
to the known result for the pure scalar case),
whereas the second term describes the effect
of the gauge-field fluctuations. Indeed, the
latter one is the dominating term and we find
that $\eta_\star$ \gl{etaf1} is negative
\beq
\eta_\star = - \frac{0.31}{N}.
\eeq

Next we turn to the second solution given by $\beta = -2$.
Here, the situation changes significantly
since  eq.~(\ref{betala}) simplifies to
\beq \label{bl2}
\frac{d \la}{d t} \!=\! -\la +
      16 e^4  v_3 l^3_2 s^3_2(2 e^2 \ka) .
\eeq
The fixed point is now given by
\beq \label{la2f}
\la_\star=16 e^4_\star  v_3 l^3_2 s^3_2(2 e^2_\star
\ka_\star)=\frac{9 l^3_2}{v_3(l^3_{gc})^2}s^3_2
(2e^2_\star \ka_\star)\frac{1}{N^2}.
\eeq
In this region, the leading contributions
to eq.~(\ref{bl2}) are coming solely from the
gauge-field fluctuations ($\sim 1/N^2$), and even
the massless scalar fluctuations contribute
only in subleading order ($\sim 1/N^3$).
The fixed point (\ref{la2f}) is thus a genuine
effect of the presence of the gauge field,
not present in the pure scalar case.
For $\ka_\star$ we find a fixed  point
$\sim N$ which corresponds to the solution of the
$N$-independent implicit equation for
$2e^2_\star \ka_\star$
\beq
2 e^2_\star \ka_\star =
8 v_3 l^3_1
e^2_\star
\left[ N +2\frac{e^2_\star}{\la_\star}
s^3_1(2 e^2_\star \ka_\star) \right] =
\frac{6 l^3_1}{l_{gc}^3}  +
\frac{l^3_{1}}{l^3_2}
\frac{s^3_1(2 e^2_\star \ka_\star)}{s^3_2(2 e^2_\star \ka_\star)}.
\label{ka2f}
\eeq
Comparing eq.~(\ref{Mf1}) with eq.~(\ref{ka2f})
we see that the numerical value for $\ka_\star$
is affected by corrections from the
gauge-field fluctuations. (Remember that the
fixed point \es \gl{e2f} is the same for both solutions
\gl{BB1}.)
We wish to put emphasis on the fact that this fixed point is
strictly different from the Gaussian fixed point even in the
limit $N \rightarrow \infty$.
Analyzing the stability properties of the fixed point
specified by (\ref{e2f}), (\ref{la2f}) and
(\ref{ka2f}) reveals two unstable
directions ($\ka$ and $\la$). We associate this point with
the tricritical fixed point.

In conclusion, we have found in the
large-$N$ approximation two non-trivial
fixed point solutions of eqs.~(\ref{betae2})-(\ref{betala}).
One solution
governs the second-order phase transition in this model.
The second fixed point is mainly an
effect due to the gauge-field fluctuations
and governs the tricritical behaviour.
Starting, for example, with $\ka_{\La}$
 near $\ka_c(\La)$,
$e^2_{\La}
= e^2_\star(\La)$
but $ \la(\La) < \la_\star(\La)$
(cf.~\gl{la2f}) the model undergoes a first-order phase transition
as
$\ka(\La)$ passes through
$\ka_c(\La)$.
Additional fixed points occur for
$e^2_\star = 0$: There is the infra-red
unstable Gaussian fixed point
at $\la_\star = 0$ with
$\ka_\star$ given by
\gl{ka1f}
and the Wilson-Fisher fixed point of the
pure scalar theory with $\la_\star$
and $\ka_\star$ given by
(\ref{la1f}) and  (\ref{ka1f}).
Since $e^2_\star = 0$ this fixed point is
different from the second-order fixed point
and has two unstable directions.
Finally, if we consider $1/e^4$ instead of $e^2$ two more
similar fixed points exist for
$(1/e^4)_\star = 0$.
Knowledge of all the fixed points
and their stability properties
permits easily to establish the qualitative
features of the whole phase diagram.\\
In section 6 we will give a detailed
numerical analysis of the second-order fixed point
and the tricritical fixed point
for arbitrary $N$. In the same section we will also
extend the analysis beyond the very simple approximation
(\ref{betae2})-(\ref{betala}) and discuss more elaborate
approximations to the flow equation \gl{exact}.

\section{ $N$-dependence of the fixed points}

For large $N$ we may obtain the fixed points of the theory
by just looking at the leading $N$-dependent
contributions
in the beta functions.
The $N$-dependence of the location of the fixed points
can be discussed in terms of a simple differential
equation which involves the derivatives of the $\beta$-functions
at the fixed point. A solution of this differential equation,
with initial conditions set at large $N$, allows to compute the
fixed points also for small values of $N$. In order to derive
the flow equation for the $N$-dependence
of the fixed points, we define the generalized vectors
\be
\vec{\la} =
\left (
\begin{array}{c}
\ka \\
\la \\
e^2
\end{array}
\right), \; \; \; \; \;
\pt \vlN = \vec{\beta}(\vlN,N) =
\left (
\begin{array}{c}
\partial_t \ka \\
\partial_t \la \\
\partial_t e^2
\end{array}
\right). \; \; \;
\ee
(This discussion can be easily extended if
additional couplings are taken into account).
In general, $\vec{\la}$ is a function
of $k$ and $N$. However, since we are
interested in the scaling solutions for arbitrary
$N$, we wish to take  $\vec{\la}$ on a
fixed point. Therefore, every scale dependence is
removed and $\vec{\la}$  becomes
solely a function of $N$,  $\vec{\la}_\star(N)$.
For all $N$ it obeys
the fixed point condition
\beq \label{fpc}
\partial_t \vec{\la}_\star(N)= \vec{\beta}(\vec{\la}_\star(N),N) = 0.
\eeq
Taking the total derivative of eq.~(\ref{fpc})
with respect to $N$ gives the desired differential equation
for the $N$-dependence of $\vec{\la}_\star$:
\beq \label{ndgl}
\frac{\partial \vec{\la}_{\star}}{\partial N}(N)
= - A^{-1}( \vec{\la}_{\star}(N),N)
\frac{\partial \vec{\beta}}
{\partial N}
(\vec{\la}_{\star}(N),N).
\label{flo}
\eeq
Here, the matrix $A(\vec{\la}(N),N)$ is
the Hessian matrix at the fixed point,
\beq
A(\vec{\la},N)=\frac{\partial \vec{\beta}}{\partial \vec{\la}}
(\vec{\la},N).
\eeq
For given $N$,
it governs the evolution of linearized fluctuations
around the scaling solution
\beq
\pt (\vec{\la} - \vec{\la}_\star)
 = A (\vec{\la} - \vec{\la}_\star)\ .
\eeq
As long as $A$ remains regular, the flow equation (\ref{flo})
has a solution and the fixed point exists
for all $N$ in this domain.
Also the stability property of the fixed point cannot
change in the domain where $A$ remains regular.
We have plotted in fig.~2 the eigenvalues of $A$
for the second-order fixed point in dependence on $N$.
All eigenvalues remain far away from zero and
depend only moderately on $N$.
Within our truncation we conclude that
a second-order fixed point and the associated
parameter region for a second-order phase transition exists for
arbitrary $N \geq 1$.
(This can formally also be continued to
$N < 1$).
We have numerically solved the flow equation (\ref{flo}),
starting initially with large $N$ and extrapolating to $N=1$.
The results that we obtained using
this method are in very good agreement with
those of the complete numerical analysis that we
present in the following section. The difference
between the two methods is indistinguishable in the plots.

\section{Numerical Analysis}
In this section we present in detail
the results of a numerical investigation of
the fixed-point structure of the abelian Higgs model.
Since we are especially interested in the critical behaviour
of the theory, we will
focus on the existence of fixed point solutions and the
related critical indices.
As we have mentioned, we employ an expansion around the
``asymmetric minimum'' at $\ka\neq 0$, thereby transforming the
partial differential equation in an infinite
series of coupled ordinary differential equations
for the corresponding couplings. This system, then,
has to be truncated at some finite order and
will be  solved numerically.

In lowest approximation,
the differential equations are given
by eqs.~(\ref{betae2}) -- (\ref{betala}).
The numerical values at the second-order  fixed point
of \ks, \ls and \es are given as functions of the number
of complex scalar fields $N$.
Our results are depicted by diamonds in figs.~3-5.
The solid lines correspond to the large-$N$
extrapolation as given by eqs.~(\ref{e2f}),
(\ref{ka1f}) and (\ref{la1f}). It turns out that \ks is  very well
described by eq.~(\ref{ka1f}) even for $N=1$.
This is related to the fact that the
contributions from the massive fluctuations in
eq.~(\ref{betaka}) are damped through
the threshold behaviour. The subleading
term at $N=1$ stems from the third term
in \gl{betaka} and is suppressed by a
factor of approximately $3 s^3_1(2  \lambda_\star \kappa_\star)
\sim
{\cal O}(10^{-2})$ compared to the leading one.
The contribution of the gauge-field fluctuation in
\gl{betaka} remains unimportant despite the
rather large value for the fixed point of
the gauge coupling (\es $\sim 633$ for $N=1$).
It is suppressed by a factor of
$4 s^3_1(2e^2_\star\ka_\star) {e^2_\star}/{\la_\star}
\sim
{\cal O}(10^{-3})$ compared to the leading term.
Fig.~4 displays the results for \ls. Again,
we find that \ls is very well
described by its large-$N$ extrapolation eq.~(\ref{la1f}).
Only for $N=1$ we observe a sizeable correction
due to the third term in \gl{betala} which
becomes nearly of the same order as the
last one. Their ratio reads $9s^3_2(2\la_\star\ka_\star)
\sim{\cal O}(10^{-1})$ and small corrections
are of no surprise. The contribution of
the gauge-field fluctuation in
\gl{betala} is suppressed by a factor of
$8 s^3_2(2e^2_\star\ka_\star) {e^4_\star}/{\la^2_\star}
\sim
{\cal O}(10^{-3})$ and therefore again
unimportant. The results for \es are given in fig.~5.
The large-$N$ extrapolation \gl{e2f}
is in a very good agreement with our numerical
results down to $N \sim 10$. For $N<10$,
approximating all fluctuations as being massless
becomes less and less accurate.
The mass of the fluctuations will lower the
$e^4$-coefficient in \gl{betae2} as
compared to the purely massless estimate
and therefore enhance the numerical
value for \es. (Clearly, this effect is largest
for $N=1$ where all contributions
in \gl{betae2}
are suppressed by massterms
and the estimate \gl{e2f} becomes
too small by roughly a factor of $10$.)

In order to check the stability of these
findings with respect to a change in the truncation, we studied
approximations of the effective potential being
a local polynomial in $(\rt-\ka)$ up to
$(\rt-\ka)^3$ and $(\rt-\ka)^4$.
(We call these approximations the
$\varphi^6$- and $\varphi^8$-approximation, respectively).
This approach turned out to be very
successful in the pure scalar model
\cite{ch2}, where the non-trivial
scaling solution corresponds to the Wilson-Fisher  fixed point.
However, we do not expect that this approach
will give reliable results near the tricritical
fixed point. Quite generally, the local
polynomial approximation around the
asymmetric minimum at $\rho_0(k)$ breaks
down when $\la(k)$ reaches zero for $k>0$
in the course of its running. This happens
near a first-order phase transition and
is connected with the fast running of $\ka$
and all the higher couplings in the region
of very small $\la$ as can be seen from
eq.~(\ref{betaka}). A discontinuity in $\ka$
(jumping from $\ka_c>0$ to $\ka=0$ at $\la=0$)
will be manifest through a pole $\sim 1/\la$
in the $\beta$--functions. Nevertheless, these regions
in parameter space can be handled by a more global approach as
described in \cite{daniel} for the $d = 4$ abelian Higgs model.
Here, flow equations at the local minimum and at the origin are used
simultaneously. At the two extrema we expand
in second-order in $\rho$ or $(\rho-\rho_0)$
and guarantee continuity of $U_k$ by assuming
a $\rho^4$ polynomial for the region between $0$
and $\rho_0$. By this method, the evolution of
$\rho_0$ knows about the flow at
the origin (for example the appearance
of a new local minimum) and vice versa.
As long as $U_k$ exhibits a local asymmetric
minimum $(\la>0,\ka>0)$, the effective
average potential is parametrized in terms of the
four variables $\la,\ka, m_s^2/k^2, $ and
$\la_s$, where $\la_s(k)=Z_{\varphi,k}^{-2}U_k''
(\rho=0)k^{-1}$ denotes
the dimensionless quartic scalar self-coupling,
and $m_s^2(k)=Z_{\varphi,k}U'_k(\rho=0)$ the scalar
mass term at the origin. This second method
will serve as an alternative check of the
results from the local polynomial approximation
at the second-order  fixed point in the
various approximations. In addition, it will
yield results for the tricritical   fixed point.

It turns out that
\ks is very stable against changes in the truncation.
The results for the $\varphi^6$- and
$\varphi^8$-approximation  and those
obtained with the global method are essentially
identical to fig.~3 and differ only
slightly from them for small $N$. The fixed points for \es and \ls
within the different approximations are depicted in fig.~6.
It shows \es as a function of \ls for different values of $N$.
The solid line corresponds to the large-$N$
estimate \gl{e2f} and \gl{la1f}:
\beq \label{e2la}
e^2_\star(\la_\star)=\frac{3 l^3_2}{l^3_{gc}}
\la_\star = 3.7\, \la_\star\,\, .
\eeq
All the different approximations converge
rather fast to the large-$N$ result \gl{e2la}.
However, for small values of $N$ the fixed
points are quite different. But this has to be
expected: The inclusion of a coupling $\sim\varphi^6$
(being dimensionless in $d=3$)
is for small $N$ an important effect for \ls
\cite{expo}. In turn, \es depends now strongly
on \ls since the massive excitations are important,
which results in a sizable shift of the second-order
fixed point. On the other side, we do not expect
that couplings $\sim\varphi^{10}$ will modify
the result of the $\varphi^8$-approximation in an important way. In
summary, the second-order  fixed point is confirmed
in a variety of approximations to the exact
evolution equation. The estimates eqs.~(\ref{e2f}),
(\ref{ka1f}) and (\ref{la1f}) give very good
predictions already for small $N$. The results
for the critical exponent $\nu$ and the anomalous
dimension $\eta$ and their discussion is given in section 7.

Finally, we turn to the tricritical fixed point.
In the $\varphi^4$-approximation we find a tricritical
point for all $N \geq 2$.
For $N=1$ the gauge coupling runs to infinity
in the interesting region and the
tricritical point presumably corresponds
to $e^2_\star \rightarrow \infty$.
This may well be an artefact of our truncation.
Our results obtained by the global approach
and the $\varphi^8$-approximation are
illustrated in fig.~7. Again, \es is given as a
function of \ls for various values of $N$.
The solid line represents the expected behaviour for large $N$
as given by \gl{e2f} and \gl{la2f}:
\beq \label{e2latri}
e^2_\star(\la_\star) \sim \sqrt{\la_\star}.
\eeq
This is well reproduced by the global method.
Here the results from the $\varphi^8$-approximation
converge rather slow to the asymptotic behaviour
(\ref{e2latri}).
A comment on the solutions
for small $N$ is in order.
For $N<4$ no tricritical fixed point has been found with
the global method. This is related to the fact
that we have to expand the potential at $\rho=0$
around a local maximum. The running of the
dimensionless mass term at the origin
\beq \label{ms}
\partial_t \, \tilde{m}^2_s = -2 \,
\tilde{m}^2_s -8v_3l^3_1 e^2 +{\cal O}
(\la_s,\eta_\varphi \tilde{m}^2_s)
\eeq
is dominated for small $N$ by the $e^2$-term.
The fixed point  for $\tilde{m}^2_s$ reads therefore
\beq
(\tilde{m}^2_{s})_\star \simeq -4v_3 l^3_1 e^2_\star.
\eeq
Requiring $(\tilde{m}^2_s)_{\star}$ to stay away
from the pole of the threshold functions at
$\tilde{m}^2_{s}=-1$ yields with the help of \gl{e2f} the condition
\beq \label{bound}
N \grgl  \frac{3 l^3_1}{l^3_{gc}} = 3.2\,\, ,
\eeq
which explains roughly the absence of fixed
points for $N \klgl 4$ in this approach.
However, we do not expect this to be a physical
effect. Instead, it is related to the fact that we
used in the global approach the evolution
equation (\ref{betae2}) not only at the asymmetric
minimum, but also at the origin. This approximation is harmless
in the large-$N$ limit where \es (\gl{e2f}) and the
fixed point for the dimensionless renormalized gauge
coupling at the origin $e^2_{0\star}$ are both small.
Their difference behaves like $1/N$. For small $N$, in
contrast, this may result in a sizeable effect. As
a result, $e^2_{0\star}$ is always smaller than
\es and the bound \gl{bound} may therefore
be lowered easily.\footnote
{In order to make a full account of this
effect, the $\rho$-dependence of $Z_F(\rho)$ (replacing
$Z_F$ in eq. (\ref{trun})) has to
be calculated. We refer this investigation to future work.}

In conclusion, it has been shown that the
second-order and tricritical fixed points are confirmed
numerically in a variety of different
approximations, locally and globally, to the exact
evolution equation (\ref{exact}). The second-order
fixed point is shown to be rather stable against higher truncations,
especially for large $N$. For large $N$ we expect a rather fast
converge of the series of approximations $\varphi^4$,
\mbox{$\varphi^6$, $\varphi^8$ ...}
towards the scaling
solution of the partial differential equation
(\ref{exact}) as given by
$\partial_t u_k (\tilde{\rho}) = 0$.
Since all contributions of terms neglected in the truncation
(\ref{trun}) are suppressed by powers of $1/N$, the predictions
of critical exponents computed from this fixed point should be
very reliable. For small $N$ the convergence
of the series of truncations is not obvious
with the restricted Ansatz (\ref{trun}).
The reason leading to the ``missing
tricritical point'' in the global
method will also affect the existence
of a scaling solution of eq. (\ref{exact})
and therefore the convergence of the series.
For small $N$ it is not excluded that the $\varphi^4$-approximation
will give better critical exponents than higher
truncations. For the tricritical point the region of convergent
truncations is restricted to even higher values of $N$.

\section{Critical exponents}

The behaviour of a system
near the critical temperature of
a second-order phase transition
is described by the
critical exponents of the theory
\cite{Amit}.
These specify
the long range correlations
of physical quantities as we approach
a zero mass theory at $T = T_c$.
The system is described by five exponents,
which however are not independent. For example
the indices $\gamma$ and $\delta$,
which describe the
response of the vacuum expectation value
$\rho_{0}$
to an external magnetic field or source,
as well as $\beta$,
which specifies
how the vacuum expectation value
depends on the temperature, may be calculated, once
both temperature dependence of
the correlation length
and the behaviour of the connected
two-point function
at the critical temperature
are known. Thus, we will look
how the
two independent indices $\nu$
and $\eta$ behave as
$N$ varies.
How these can be computed from
properties of the
scaling solution
is the subject of this section.

The critical exponent
$\eta$
determines
the behaviour of the connected two-point
function at the critical
temperature $(T=T_{c})$.
\beq
\lim_{x \rightarrow \infty}
< \varphi(x) \varphi(0)>_{c}
\sim |x|^{-(d-2+\eta)} .
\label{inxi}
\eeq
It arises from the anomalous dimension
of the field $\varphi$
and is directly related to the momentum
dependence of the
(unrenormalized) Green function for
$q^2 \rightarrow 0$ at $T = T_c$
\beq \label{g-1}
G^{-1}(q) = q^2 Z(q)
\eeq
\beq
\lim_{q^2 \rightarrow 0} Z(q) \sim (q^2)^{-\frac{\eta}{2}} .
\eeq
The Green function $G^{-1}(q)$ is encoded in the
momentum dependence
of the derivative terms in $\Gamma_k$
which are quadratic in small fluctuations around the
$\varphi = const$ configuration.
Its computation goes beyond the
truncation (\ref{trun}) and
would necessitate a generalization
from a constant $Z_{\varphi,k}$ to a momentum
dependent function
$Z_k(q)$. Then $Z(q)$ in \gl{g-1} obtains as
$\lim_{k \rightarrow 0} Z_k(q)$
whereas
$Z_{\varphi,k}$
is given by
$\lim_{q^2 \rightarrow 0} Z_k(q)$.
Even without computing
$Z_{k}(q)$
explicitly, simple scaling considerations
indicate for the critical behaviour
\beq
Z_k(q) = f \left (
\frac{q^2}{k^2} \right )
\left (
\frac{q^2+k^2}{\Lambda^2}
\right )
^{-\frac{\eta}{2}} .
\eeq
Here $\La$ is some appropriate high momentum
scale and the smooth function $f$
approaches constant values for both limits
$q^2 \ll k^2$ and
$q^2 \gg k^2$. This follows from the observation that
$q^2$ and $k^2$ act as independent
infra-red cut-offs in the loop describing
the exact flow equation (\ref{1.20}).
In consequence, the critical exponent $\eta$
can be extracted from the
$k$-dependence
of the wave function renormalisation
for the scaling solution
\beq
\eta =  - \left (
\frac{\partial}{\partial t}
\ln Z_{\varphi, k} \right)_{\star} .
\eeq
This allows to identify $\eta$
with $\eta_{\varphi}$ (\ref{andim}),
where all couplings are taken at their fixed point values.

We now turn to the critical index
$\nu$,
which is related to the temperature
dependence
of the correlation length
near the critical temperature
$T_{c}$
\beq
\lim_{x \rightarrow \infty}
<\varphi(x)\varphi(0)>_{c}\
\sim e^{-x/\xi}, \; \; \;
\xi = m^{-1} \sim |T-T_{c}|^{-\nu} .
\label{ntc}
\eeq
Here the
renormalized mass $m^2$
is given by the
limit $k \rightarrow 0$
of the $k$-dependent mass term
$m^2(k) = Z_{\varphi,k}^{-1}
(k) U'_{k}(0)$
and $2 Z_{\varphi,k}^{-1}
U''_{k}(\rho_{0}(k))
\rho_{0}(k) = 2 \la(k)\ka(k)k^{2}$,
for the symmetric phase
and the phase with spontaneous symmetry breaking (SSB),
respectively.
The phase transition
corresponds to a critical
trajectory
within the SSB regime
$m^{2}_{\star}(k) =
2 \la_{\star}\ka_{\star}
k^{2}$.\\
In order to extract $\nu$
we study a small deviation
from the critical trajectory
\beq
m^{2}(k) =  m^{2}_{\star}(k) +
\delta m^{2}(k) .
\eeq
The evolution equation for
$\delta m^{2}$
is characterized by the anomalous mass dimension
$\omega$
\beq
\frac{\partial}{\partial t}
\delta m^{2} = \omega \; \delta m^{2} .
\label{eqlin}
\eeq
For very small
$\delta m^{2}/k^{2}$
we linearize in
$\delta m^{2}$ so that $\omega$
becomes independent of
$\delta m^{2}$
\beq
\omega = \frac{\partial \beta_{m^{2}}}
{\partial m^{2}} (m^{2}=m_{\star}^{2})
\eeq
\beq \label{w}
\beta_{m^{2}} = \frac{\partial m^{2}}
{\partial t}
= (2 + \frac{\bl}{\la}
+ \frac{\bk}{\ka})m^{2} .
\eeq
At the fixed point
the anomalous mass dimension is a constant
$ \omega_{\star} = \omega_{\star}
(\ka_{\star},\la_{\star},e^2_{\star})$.
We therefore obtain in the
vicinity of the scaling solution
for $\delta m^{2} \ll k^{2}$
\beq
\delta m^{2}(k) =
\left ( \frac{k}{k_{1}} \right)^{\omega_{\star}}
\delta m^{2} (k_{1}) .
\label{E1}
\eeq
(Here $k_{1}$ denotes
some scale below the
ultra-violet cut-off
for which all couplings are
near their fixed points.
It can be taken proportional to
$T_{c}$).
For
$\omega < 2$
the ratio
$\delta m^{2} / k^{2}$
increases
as $k$ becomes smaller.
For any non-vanishing
$\delta m^{2}(k_{1})$
there is necessarily
a critical scale
$k_{c}$
where the linearization
of (\ref{eqlin})
becomes invalid,
typically for some constant
$c < 1$
\beq
\delta m_{c}^{2} = \delta
m^{2}(k_{c}) = c k_{c}^{2} .
\label{E2}
\eeq
Near the phase transition
the renormalized
mass is the only
scale present. In the
SSB phase
we conclude
from dimensional analysis
that it must be proportional to
$k_{c}$
\beq
m^{2} = m^{2}(0) =
\delta m^{2}(0) = a k_{c}^{2} .
\label{L1}
\eeq
The situation is similar in the symmetric
phase, but the
proportionality constant
$a$ is in general
different
for the
symmetric and
the SSB phase. We
finally observe that
$m^{2}(k_{1})$
is a function of temperature
and that in linear
approximation
\beq
m^{2}(k_{1},T_{c}) = m^{2}_{\star}(k_{1}),
\; \; \; \;
\delta m^{2}(k_{1}) = b (T_{c}-T) .
\label{E3}
\eeq
(Remember that the symmetric phase
corresponds to
$
m^{2}(k_{1}) < m^{2}_{\star}(k_{1})
$).
Combining
eqs.~(\ref{E1}),(\ref{E2}) and (\ref{E3}) gives
\beq
k_{c}^{2} = \frac{b}{c}
\left (
\frac{k_{c}}{k_{1}}
\right )^{\omega_{\star}}
(T_{c}-T) ,
\eeq
and we conclude
\beq
m^{2} = a
\left |
\frac{b}{c}
k_{1}^{-\omega_{\star}}
\right |
^{\frac{2}{2-\omega_{\star}}}
|T-T_{c}|
^{\frac{2}{2-\omega_{\star}}} .
\eeq
This is to be compared with
eq.~(\ref{ntc}),
showing that the critical exponent $\nu$
is directly related to
$\omega_{\star}$
\beq
\nu = \frac{1}{2-\omega_{\star}} .
\label{nuex}
\eeq
We emphasize that
(\ref{inxi})
with (\ref{nuex}) becomes exact
in the limit $T \rightarrow T_{c}$
since the fixed point behaviour
completely determines
the evolution
of $m^{2}$. In other words,
the
``initial'' and ``final''
running of $m^{2}$
away from the fixed point becomes
negligible
compared to the running near the fixed point.
The preceding discussion has been explicitely
verified by numerical studies for the pure
scalar theory \cite{ch2}.\\
We still have to specify how the
anomalous mass dimension
$\omega_{\star}$
should be evaluated.
In fact the ratios $\s0{\bl}{\la}$ and $\s0{\bk}{\ka}$
in \gl{w}  depend not only on
$\la \ka$ or $m^2$ which we have up to now taken as
a variable for simplicity of the presentation.They also involve
 $\la$ and
$e^2$  (and in principle also some  other parameters
characterizing the average action,
like
for example the $\varphi^{6}$
coupling). For a computation of $\omega_{\star}$
we have to specify how the
couplings we are using depend on each other.
Instead of $m^2$ we use here $\ka$
as the independent parameter and express
$\la$ and $e^2$ as a function of $\ka$.
(This can easily be extended if additional
couplings are considered). Then $\omega_{\star}$
can be expressed in terms of the partial derivatives
of $\beta_{\ka}$ evaluated at the fixed point:
\beq
\omega_{\star}
= 2 + \left (
\frac{\partial \bk}{\partial \ka}
\right )_{\star}
+
\frac{\partial \la}{\partial \ka}
\left (
\frac{\partial \bk}{\partial \la}
\right )_{\star}
+
\frac{\partial e^2}{\partial \ka}
\left ( \frac{\partial \bk}{\partial e^2}
\right )_{\star} .
\eeq
It remains a determination of the
quantities
$ \frac{\partial \la (\ka)}{\partial \ka}$
and
$ \frac{\partial e^2(\ka)}{\partial \ka}$.
The phase transition $(T = T_{c})$
corresponds to a running along a trajectory on the critical surface
  into the second-order fixed point.
Trajectories infinitesimally
close to the critical surface
(for $T$ infinitesimally
close to $T_{c}$)
come infinitesimally close
to the fixed point
and then move away from it infinitesimally
close to
a curve denoted as
``unstable direction''.
The dependence $\la(\ka)$ and $e^2(\ka)$
on the curve corresponding to
the unstable direction should be used
in the equation for
$\omega_\star$. More formally,
the evolution equation for
small deviations from
the fixed point
\beq
\delta \ka
= \ka - \ka_{\star},
\; \;
\delta \la
= \la - \la_{\star},
\; \; \delta e^2 =
e^2 - e^2_{\star}
\eeq
can be written as
\beq
\frac{\partial}{\partial t}
\left(
\begin{array}{c}
\delta \ka
\\
\delta \la
\\
\delta e^2
\\
\end{array}
\right) =
A
\left(
\begin{array}{c}
\delta \ka
\\
\delta \la
\\
\delta e^2
\\
\end{array}
\right)
\eeq
with (cf.\,\,sect.\,4)
\beq
A =
\left(
\begin{array}{ccc}
\partial \bk / \partial \ka
& \partial \bk / \partial \la
& \partial \bk / \partial e^2
\\
\partial \bl / \partial \ka
& \partial \bl / \partial \la
& \partial \bl / \partial e^2
\\
\partial \bee2 / \partial \ka
& \partial \bee2 / \partial \la
& \partial \bee2 / \partial e^2
\\
\end{array}
\right) .
\eeq
At the second-order fixed point
all eigenvalues of the matrix $A$,
except one, are positive.
Let us denote by
$x^{us} = (1, b_2, b_3)$
the eigenvector corresponding to
the negative eigenvalue.
As $k$ decreases,
only this eigenvector plays a role.
It indicates the unstable direction
whereas the components corresponding to
positive eigenvalues quickly die out.
For the unstable direction
we therefore find that
$ \frac{\la}{\ka}
= b_{2}$,
or, for the equation for
$\omega_{\star}$,
$ \frac{\partial \la}{
\partial \ka}
= b_{2}$,
$ \frac{\partial e^2}{
\partial \ka}
= b_{3}$.
Since in general $A$ is not
a symmetric matrix, we should specify what
we mean by an eigenvector.
We denote by $\theta$
the unique, negative eigenvalue of $A$.
Using the transformation
$ \tilde{A} = C^{-1}AC $
we can bring $\tilde{A}$
to a form where
$ \tilde{A}_{i1} = \theta \delta_{i1}$.
Then the unstable direction corresponds
to the vector
\beq
\tilde{x}^{us}_{i} =
C^{-1}_{ij} x_{j}^{us} = c \delta_{i1}
\label{tilxi}
\eeq
with
\beq
\frac{\partial }{\partial t}
\tilde{x}^{us} = \theta \tilde{x}^{us} .
\eeq
Inverting (\ref{tilxi})
we can read off
$ b_{i} = \frac{C_{i1}}{C_{11}}$
where the coefficients $C_{i1}$
obey
$ A_{ij}C_{j1} = \theta C_{i1}$.
This completely specifies $\omega_{\star}$
by the derivatives of the
$\beta$-functions at
the fixed point.
Knowledge of the fixed point values of the couplings
for arbitrary $N$ and the corresponding
$\beta$-functions provides the value of $\nu$
by simple algebraic manipulations.

\section{Results and comparison with
$\epsilon$-expansions}

In this section, we compare our results for the
 critical indices at the
 second-order phase transition
with those obtained earlier by other authors
for large values of $N$.
Our numerical results are given in figs.~8 and 9
 for the anomalous dimension $\eta$ and the critical
 index $\nu$, respectively.
Different methods have been tried in the past
to compute these critical exponents
for large $N$ \cite{suco,hikami}. Since
all methods (including ours)
are thought to give the most accurate values for
large $N$, a comparison of the results
is best made for $N \rightarrow \infty$
where $\eta \sim \frac{1}{N}$
and $\nu-1 \sim \frac{1}{N}$. In table 1 we compare
the coefficients $N \eta(N)$ and
$N(\nu(N)-1)$ as obtained by various methods. In our approach the
different approximations (and the global method)
converge very well for these quantities
in the large-$N$ limit. Our result of the anomalous dimension
is given by eq.~(\ref{etaf1}).
We have not computed an analytical expression
for the subleading coefficient
of $\nu$. This would require a systematic expansion
of the quantities appearing in the
last section in powers of $1/N$.
However, this coefficient has been obtained by a numerical fit of the
data for $\nu(N)$ to a polynomial in $1/N$ and we find $ \nu = 1
-\frac{1.38}{N}$.

Two papers quote results of a large-$N$ estimate from a computation
in fixed dimension $d=3$. We should emphasize here that
the situation for a gauge theory is rather different from
the pure $2N$-component scalar theory where exact large-$N$
results can be obtained. A loop computation
in three dimensions requires the presence of an effective
infra-red cut-off for all fields. In the pure scalar
theory this is provided by the scalar mass term if the phase
transition is approached
from the symmetric phase
\cite{Parisi}.
For the gauge model investigated here the photon remains
massless in the symmetric phase and produces an
infra-red divergence in the loop expansion. This problem
persists in the large-$N$ approximation.
Thus perturbative calculations in three dimensions
have to introduce some infra-red cut-off
by hand. We give the estimate of such
methods \cite{suco,hikami}
in table 1. In our approach, the infra-red cut-off is always
provided by the averaging scale $k$ and no additional assumptions
have to be made. We find a substantial discrepancy of previous
large-$N$ estimates from our result for $d=3$.
Nevertheless, we expect that the relatively crude estimates
of ref. \cite{suco,hikami}
give rather realistic results for $d \approx 4$ and $d \approx 2$.
For $d=4$ the infra-red divergence is only logarithmic
and the exact form of the infra-red cut-off
does not matter.
For $d=2$ there are no transverse photon degrees of freedom and the
infra-red problem in the loop expansion may be
circumvented.

\begin{table}
\centering
\begin{tabular}
{|l|c|c|}
\hline
 The limit $N\rightarrow \infty$ of& $N\eta(N)$ & $N[\nu(N)-1]$  \\
 \hline \hline
Our result: & $-0.31$ & $-1.38 $ \\
Previous large-$N$ estimates \cite{suco,hikami}:
& $-2.03$ & $-4.86$ \\
$\epsilon$-expansion at $\epsilon=1$ around&&\\
\quad a) $d=4-\epsilon$ \cite{suco}: &$-9 +
{\cal O}(\epsilon^{2})$&$-48 + {\cal O}(\epsilon^{2})$\\
\quad b) $d=2+\epsilon$,
$CP^{N-1}$ model, order $\epsilon^2$
\cite{hikami}:& & $ -2+ {\cal O}(\epsilon^{3})$ \\
\quad c) $d=2+\epsilon$,
$CP^{N-1}$ model, order $\epsilon^4$
\cite{hikami}:& &
$-6 + {\cal O}(\epsilon^{5})$\\
\hline
\end{tabular}
\begin{center}
{\bf Table 1}
\end{center}
\label{table1}
\end{table}

As an alternative to fixed dimension large-$N$ estimates
one may attempt to use
an extrapolation of results in $4-\epsilon$ dimensions
to $\epsilon = 1$ ($\epsilon$-expansion) \cite{suco,LN}.
Results of this method for
the critical exponents
are also shown in table 1. We emphasize
that the result in order $\epsilon$ differs
from our result by factors of about $30$ for
both $N \eta$ and $N(\nu-1)$! These rather large differences
can directly be traced to the threshold effects for the massive
fluctuations which are neglected in lowest order $\epsilon$-expansion.
This can be seen by neglecting the masses in eq. (\ref{etaf1}):
The leading $1/N$ coefficient in the anomalous dimension would then
reach $-9.53$ and be of the same order as
the $\epsilon$-expansion result
$-9$. For $d \rightarrow 4$ $(\epsilon \rightarrow 0)$ the mass terms
are small at the fixed point and the $\epsilon$-expansion
becomes reliable. For $d=3$, however, the mass terms
are substantial and the $\epsilon$-expansion fails
to reproduce the leading $1/N$ coefficient correctly.
This is directly related to the absence of a second-order fixed point
for moderate values of $N$ within the lowest order
$\epsilon$-approximation. In contrast to the $O(N)$ symmetric pure
scalar theory, where the $\epsilon$-expansion
gives very good estimates for the critical exponents \cite{scalar},
we conclude that the $\epsilon$-expansion gives a very misleading
picture of the nature of the phase transition
for small values of $N$ and sufficiently large
$\lambda /e^2$.

\begin{table}
\centering
\begin{tabular}
{|c|c|c|c|c|}
\hline
$N$   & $\eta (\varphi^4)$  & $\eta (\varphi^8)$
& $\nu (\varphi^4)$ & $\nu (\varphi^8)$ \\
\hline \hline
1  & --0.134   & --0.170   & 0.532  &  0.583  \\
2  & --0.0892  & --0.123   & 0.639  &  0.568  \\
3  & --0.0672  & --0.0835  & 0.715  &  0.664  \\
5  & --0.0453  & --0.0513  & 0.798  &  0.777  \\
7  & --0.0343  & --0.0374  & 0.843  &  0.832  \\
10 & --0.0252  & --0.0267  & 0.883  &  0.877  \\
30 & --0.00909  & --0.00926  & 0.956  &  0.956  \\
100 & --0.00281 & --0.00283  & 0.986  &  0.986  \\
\hline
\end{tabular}
\begin{center}
{\bf Table 2}
\end{center}
\label{table2}
\end{table}

Finally, in the limit $e^2 \rightarrow \infty$
the $N$-component abelian Higgs model is thought to be described
by a pure scalar $CP^{N-1}$ model
\cite{CPN-1}.
We notice that towards the infra-red $e^2(k)$ decreases
very fast for large $N$. The effect of $ 1/e^2 \neq 0$
corresponds therefore to a relevant parameter
in the appropriately perturbed $CP^{N-1}$ model.
Nevertheless, if $e^2_\star \kappa_\star$ remains very large
at the true second-order fixed point
of the abelian Higgs model, the critical exponents
may be close to the ones of the $CP^{N-1}$ model.
The critical exponents of the $CP^{N-1}$ model
have been computed by an expansion around
$2 + \epsilon$ dimensions up to order $\epsilon^4$ \cite{hikami}.
We notice a rather poor convergence of the $\epsilon$-series for large
$N$ and $\epsilon =1$. The quoted values of the leading
$1/N$ coefficients are also tabulated in table 1
and turn out to be much larger than our results.
At this stage it is not clear if this is a failure
of the $\epsilon$-expansion
for the $CP^{N-1}$ model,
or if the $1/N$ coefficients
of the critical exponents in the abelian Higgs model
differ substantially from the ones of the $CP^{N-1}$ model.

Let us next turn to small values of $N$. The rather small values
of the $1/N$-coefficients of $\eta$ and $\nu-1$
allow a rather smooth extrapolation towards small $N$
(cf. figs.~8 and 9).
This is in contrast to previous computations.
In order to see the effects of the approximations we have depicted
in figs.~8 and  9
solutions of two different approximations
to \gl{exact}. For $N>10$ their difference is negligible. However, for
$N<10$, effects of higher couplings begin to have a
sizeable influence on the critical
exponents $\eta$ and $\nu$. As discussed briefly in
sect. 5, we do not expect a good convergence
of the truncation series for small $N$. This is also
apparent from our results for $N \leq 2$. The most reliable
estimate at present is probably the one
of the $\varphi^4$-approximation.

In table 2, we give the critical indices of
the second-order phase transition for
different values of $N$ and for different approximations.
We notice that for all values of $N$ the anomalous dimension
turns out negative. This is in contrast to
pure $O(N)$-symmetric scalar theories and is due to the gauge
field fluctuations. Since for $N=2$ the $CP^1$ model is equivalent
to an $SO(3)$ model which has positive $\eta$
we conclude that for small $N$ our critical exponents
differ substantially from the $CP^{N-1}$ model.
It is also clear that the critical exponents deviate
from the $O(2N)$ scalar theory which is obtained for $e^2 =0$.
This is true for all $N$.

\section{Conclusions}

We have presented a non-perturbative analysis
of the scaling behaviour of the three-dimensional abelian
Higgs model for $N$ complex scalar fields.
It is based on the investigation of fixed points
of non-perturbative flow equations
which describe the scale dependence
of the average action. One fixed point governs the second-order
phase transition (for the appropriate parameter range)
whereas the other corresponds to the tricritical
point where the second-order transition changes
to a first-order one. We have found a second-order fixed point
for all values of $N$, including the case of the superconductor
for $N=1$
\footnote{The case $N=1$ is studied in
full detail in
ref. \cite{ber}.}. This suggests a second-order
phase transition for type II superconductors
with sufficiently strong scalar coupling.
Such a picture is consistent with lattice studies
\footnote{Lattice studies cannot distinguish between
a second-order and a weakly first-order transition.} \cite{lattice}
. We have computed the critical exponents
$\eta$ and $\nu$ in dependence on $N$.
For $N=1$ we obtain $\eta$ between  $-0.13$
and $-0.17$ and $\nu$ between $0.53$ and $0.58$.
 From the scaling laws we infer $\alpha$ between $0.25$ and $0.4$,
$\beta$ between $0.23$ and $0.24$ and $\gamma$ between
$1.254$ and $1.135$.
The phase transition from nematic to smectic-A in
liquid crystals
is thought to be modeled by the same
universality class
as superconductors. Our values for the critical
exponents may therefore be compared to the measured values
for the transition in these liquid crystals.
The indices $\alpha$ and $\gamma$ agree within the
experimental uncertainties.
For $\nu$ the experiment distinguishes
between parallel and perpendicular directions
\cite{Joetal},
and direct comparison with our value is more
difficult.

We find a very smooth transition between the behaviour for
large and small $N$. In particular, the $1/N$-expansion
seems to remain valid down to rather small values
of $N$. We were also able to find the tricritical fixed
point, even though our global method could not
be used for $N < 4$. We have not attempted here to
compute the crossover exponents associated to
this fixed point. This can be done by methods in complete
analogy to sect.~6. We expect our results for the tricritical
fixed point to be less reliable than the ones
for the second-order fixed point.
This is connected to shortcomings of the present
truncation that will be discussed below.
An important message from the present work concerns the
validity of the $\epsilon$-expansion. In contrast to
$O(N)$-symmetric scalar theories we find that the
$\epsilon$-expansion gives for all $N$ a rather poor
description of the phase transition in
the three-dimensional abelian Higgs model.
For large $N$ the $\epsilon$-expansion results for
the leading $1/N$ coefficients of $\eta$ and $\nu-1$
are off by a factor of $30$.
For low $N$ the $\epsilon$-expansion
would suggest a first-order transition
for all parameter values, in contrast to our
findings.

Let us finally discuss the limits of quantitative validity
of our results. This mainly concerns
the truncation (\ref{trun})
which leads to the non-perturbative flow
equations (\ref{exact}), (\ref{andim}) and (\ref{betae2}).
This truncation can be viewed as a lowest
order approximation of a systematic derivative expansion
of the most general form of the average action
$\Gamma_k$. The next order should replace the
coefficient $Z_{F,k}$
in eq.\,(\ref{trun})
by a function
$Z_{F,k}(\rho)$
and similar for the scalar kinetic terms.
This will lead to additional contributions
to the flow equations for the potential
and the anomalous dimension $\eta_{\varphi}$
which involve $\rho$-derivatives
like $Z'$ \cite{ch1,expo}.
The most important modification, however,
is probably the effective replacement
of $e^2$ by a $\rho$-dependent function.
Furthermore, it may become necessary
to take the effective momentum dependence
of $e^2$ into account.
All this can be done by exploiting
the exact flow equation
with less severe truncations. In the present
truncation, eq.\,(\ref{trun}),
the treatment of the region around the origin
$(\rho = 0)$ seems not very reliable in the
SSB regime where the potential minimum occurs
for $\rho_0 \neq 0$. In particular, we use at the
origin effectively $e^2(\rho_0)$ instead of
$e^2(0)$. The resulting errors are particularly important
for small $N$ and we believe that this is the reason
why we did not find the tricritical fixed point
with the global method for $N < 4$. Similarly, it seems
probable that no exact scaling solution
exists for the full system of flow equations
(\ref{exact}), (\ref{andim}) and (\ref{betae2})
for small $N$
\cite{nikt}.
Since the problem appears to be
mainly related to the behaviour at the origin,
a local approximation (in $\rho$)
seems more appropriate at the moment for the
treatment of the second-order fixed point.
A definite computation of the parameter range
where the phase transition in
 superconductors is second-order requires an understanding
of the mentioned truncation problems and the
demonstration of a fully scale invariant
solution for $u_k(\tilde{\rho})$.

\pagebreak
%
%
%
\begin{center}
{\Large \bf Appendices}
\end{center}
\begin{appendix}
\section{Functional derivative of the average action}
For the derivation of the second functional
derivative $\Gamma_k^{(2)}$, evaluated at
$\varphi_a = \sqrt{\rho} \delta_{a1}$ and
arbitrary $A_\mu$ we make the following decomposition \cite{ch3}
of the fluctuations
$\delta \varphi$, $\delta A$:
\be
\d A_\m\!&=&\!t_\m + \sqrt{\alpha}
\frac{\partial_{\mu}}{\sqrt{-\partial^2}}
\tilde{\ell}\; , \; \partial^\mu t_\mu=0 \nonumber\\
\d \varphi_1\!&=&\!\frac{1}{\sqrt{2}}
\left(\sigma + i \omega\right) \nonumber\\
\d \varphi_a\!&=&\!\frac{1}{\sqrt{2}}\left(\tilde{\sigma}_{a} + i
\tilde{\omega}_{a}\right), \mbox{\ \ $a>1$} .
\label{3.2.3}
\ee
Concerning the scalar
sector we observe one massive mode $\sigma$,
one massless Goldstone mode $\omega$
and $2(N-1)$ massless pseudo-Goldstone bosons
$\tilde{\sigma}_{a}$,
$\tilde{\omega}_{a}$, $a > 1$.
Working in the Landau gauge ($\alpha=0$)
one finds
\be
\i d^dx \delta\psi\tilde{\Gamma}^{(2)}
\delta\psi = & \i d^dx \Big\{ & t_\m [
\Zf P(-\partial^2) +2 \bar{e}^2 \Zp \rho ] t^\m
\nonumber\\  &&
+\tilde{\ell}P(-\partial^2) \tilde{\ell} +
\sigma [\Zp P(-D^2)_{sy} +U'_k + 2 \rho U''_k ] \sigma
\nonumber \\ &&
+ \omega[\Zp P(-D^2)_{sy} +U'_k] \omega +
4 \sqrt{2} \bar{e}^2 \sqrt{\rho} \Zp t_\m A^\m \sigma
\nonumber \\ &&
 + 2 \Zp \omega P(-D^2)_{a s}\sigma
\nonumber\\  &&
+\sum_{a=2}^{N}\Big( \tilde{\sigma}_a[\Zp P(-D^2)_{sy} +
U'_k] \tilde{\sigma}_a
\nonumber \\ &&
+\tilde{\omega}_a[\Zp P(-D^2)_{sy}+U'_k] \tilde{\omega}_a
\nonumber \\ &&
+2 \Zp \tilde{\omega}_a P(-D^2)_{as} \tilde{\sigma}_a \Big) \Big\},
\label{actio}
\ee
where
\beq
F[A]_{sy} = \frac{1}{2}
(F[A] + F[-A]), \; \; \;
F[A]_{as} = \frac{1}{2i}
(F[A] - F[-A]) .
\eeq
Putting $A_{\mu} = 0$ $(D^2 = \partial^2)$ and
switching to momentum space, yields
$\Gamma_k^{(2)} [ \varphi, 0]$
as needed for the derivation of the flow
equation (\ref{pde}) for the average potential.

\section{Threshold functions and constants}
We use the following threshold functions and constants:
\be
l^d_0 \!&=&\! \frac{1}{2} k^{- d} \int_0^{\infty} dx
     x^{\frac{d}{2}-1}
     \tilde{\partial}_t
   \ln(P(x))\nonumber \\
l^d_{n \geq 1} \!&=&\!
     - \frac{1}{2} k^{2 n - d} \int_0^{\infty} dx
     x^{\frac{d}{2}-1}
     \tilde{\partial}_t
    P(x)^{-n} \nonumber \\
l^{d \geq 3}_g  \!&=&\!
     - \frac{d-2}{4} k^{4-d} \int_0^{\infty}
     dx x^{\frac{d}{2}-2} \frac{d}{dx} \left[
     \frac{1}{P(x)} \tilde{\partial_t} P(x)
     \right] \nonumber\\
s^d_{0}(\omega) \!&=&\!
      \frac{1}{2 l^d_0} k^{- d} \int_0^{\infty} dx
     x^{\frac{d}{2}-1}
     \tilde{\partial}_t \ln\left(
    P(x)+ \omega k^2 \right)  \nonumber \\
s^d_{n \geq 1}(\omega) \!&=&\!
     - \frac{1}{2 l^d_n} k^{2 n - d} \int_0^{\infty} dx
     x^{\frac{d}{2}-1}
     \tilde{\partial}_t \left(
    P(x)+ \omega k^2 \right) ^{-n} \nonumber \\
l^d_{n,m}(\omega_1, \omega_2) \!&=&\!
     - \frac{1}{2} k^{2 (n+m) - d} \int_0^{\infty} dx
     x^{\frac{d}{2}-1} \times \nonumber \\
     & &\! \tilde{\partial}_t
     \left[ \left( P(x) + \omega_1 k^2
          \right)^{-n}
     \left( P(x) + \omega_2 k^2
          \right)^{-m} \right] \nonumber\\
m^d_{n,m}(\omega_1, \omega_2) \!&=&\!
     - \frac{1}{2} k^{2(n+m-1)-d} \int_0^{\infty} dx
     x^{\frac{d}{2}} \times \nonumber \\
     & &\!  \tilde{\partial}_t \left[
     \left( \frac{dP}{dx} \right)^2
     \left( P(x) + \omega_1 k^2
          \right)^{-n}
     \left( P(x) + \omega_2 k^2
          \right)^{-m} \right] \nonumber\\
n^d_{n,m}(\omega_1, \omega_2) \!&=&\!
     - \frac{1}{2} k^{2(n+m-1)-d} \int_0^{\infty} dx
     x^{\frac{d}{2}} \times \nonumber \\
     & &\!  \tilde{\partial}_t \left[
     \frac{dP}{dx}
     \left( P(x) + \omega_1 k^2
          \right)^{-n}
     \left( P(x) + \omega_2 k^2
          \right)^{-m} \right] \nonumber\\
l^{d \geq 3}_c(\omega)  \!&=&\!
     \frac{d-2}{4} k^{4-d} \int_0^{\infty}
     dx x^{\frac{d}{2}-2} \frac{d}{dx} \frac{d}{dt}
     \ln \left( 1 + \frac{P(x)-x}{k^2(1+\omega)}
     \right) \nonumber \\
l^d_{gc}\!&=&\!l^d_g +l^d_c(0) .
\label{B.1}
\end{eqnarray}
Here
$\tilde{\partial}_t$
stands for
$\frac{\partial}{\partial t}$
acting only on $R_k$ contained implicitely in $P(x)$  (see sect.~2).
These functions, as well as
expressions for the asymptotic behaviour
of $s^d_n(\omega)$
for $\omega \rightarrow \infty$
can be found in
\cite{expo}. In terms of the above,
the threshold function from the massive
fluctuations for   $\bee2$ in \gl{betae2} reads
\be
\tilde{s}^d_g(2 \la \ka,
2 e^2 \ka ) \!&=&\! \frac{48 e^2 \ka}{d (d+2) l^d_g} \left[
(d+1) m^d_{2,2}(2 \la \ka, 2 e^2 \ka) -
n^{d-2}_{2,1}(2 \la \ka, 2 e^2 \ka) \right] \nonumber \\
& &\! +\frac{m^{d+2}_{2,2}(2 \la \ka, 0)}{m^{d+2}_{2,2}(0,0)}.
\label{B.2}
\ee
Furthermore,
for large mass terms $ \omega_2 >> 1$
we have
\be
\lim_{\omega_2 \rightarrow \infty}
l^d_{n,m}(\omega_1,\omega_2) \!&=&\!
l^d_n s^d_n(\omega_1) \omega_2^{-m} + {\cal{O}}(\omega_2^{-(m+1)})
\nonumber \\
\lim_{\omega_2 \rightarrow \infty}
m^d_{n,m}(\omega_1,\omega_2) \!&=&\!
m^d_{n,0}(\omega_1, 0) \omega_2^{-m} + {\cal{O}}(\omega_2^{-(m+1)})
\nonumber \\
\lim_{\omega_2 \rightarrow \infty}
n^d_{n,m}(\omega_1,\omega_2) \!&=&\!
n^d_{n,0}(\omega_1,0 ) \omega_2^{-m} + {\cal{O}}(\omega_2^{-(m+1)})  .
\label{B.3}
\ee
\end{appendix}

\newpage

\section*{Figure captions}
\begin{enumerate}
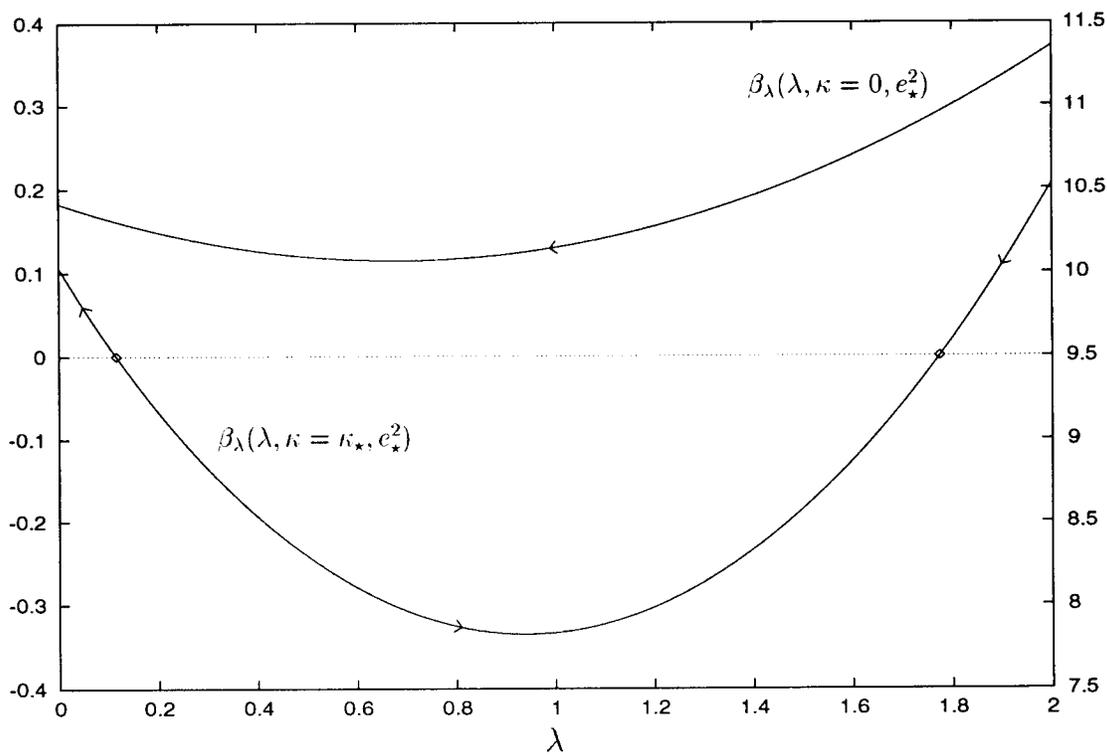
\item[{ Fig.~1:}]  The $\bl$-function \gl{betala} is
 given in dependence on $\lambda$ for
 $N=10$ and two values of $\kappa$. The arrows indicate
the flow with scale $k\rightarrow
 0$. No fixed points are obtained when the threshold behaviour
  is discarded, $\ka=0$ (upper curve, right
 axis). In contrast, the inclusion of the threshold
 effects with $\ka=\ka_\star$
allows for two fixed points (lower curve, left axis).
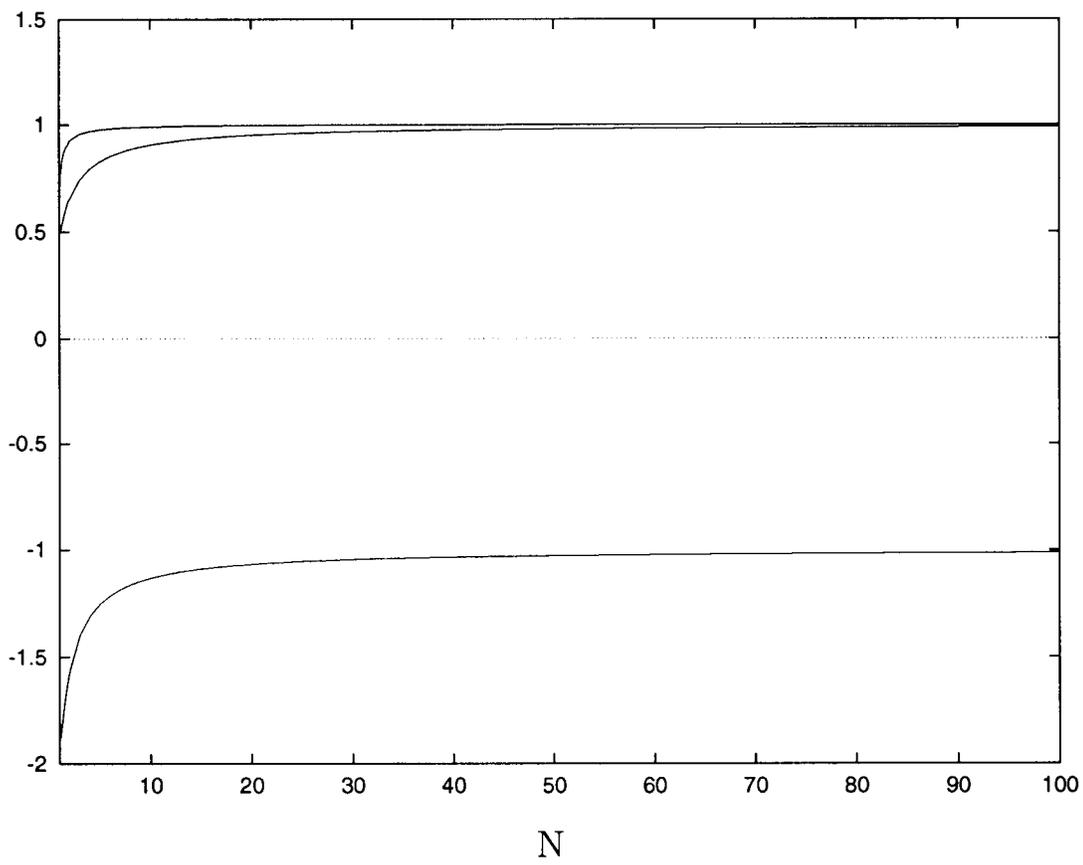
\item[{ Fig.~2:}]  Eigenvalues of the fluctuation matrix $A$
in dependence on $N$. The two positive eigenvalues
indicate that the fixed point corresponds
to a second-order phase transition
for all $N$.
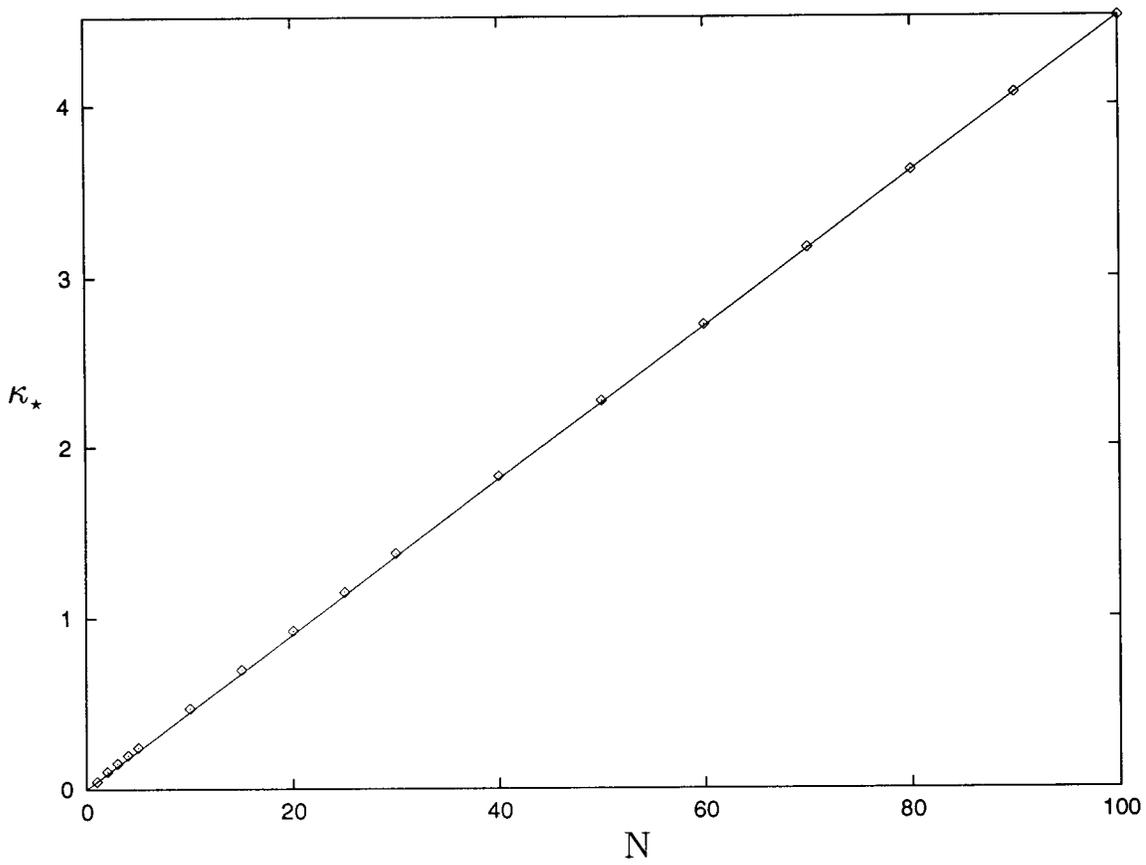
\item[{ Fig.~3:}]  Fixed point value $\kappa_\star$ in dependence on
$N$ for a $\varphi^4$-approximation.
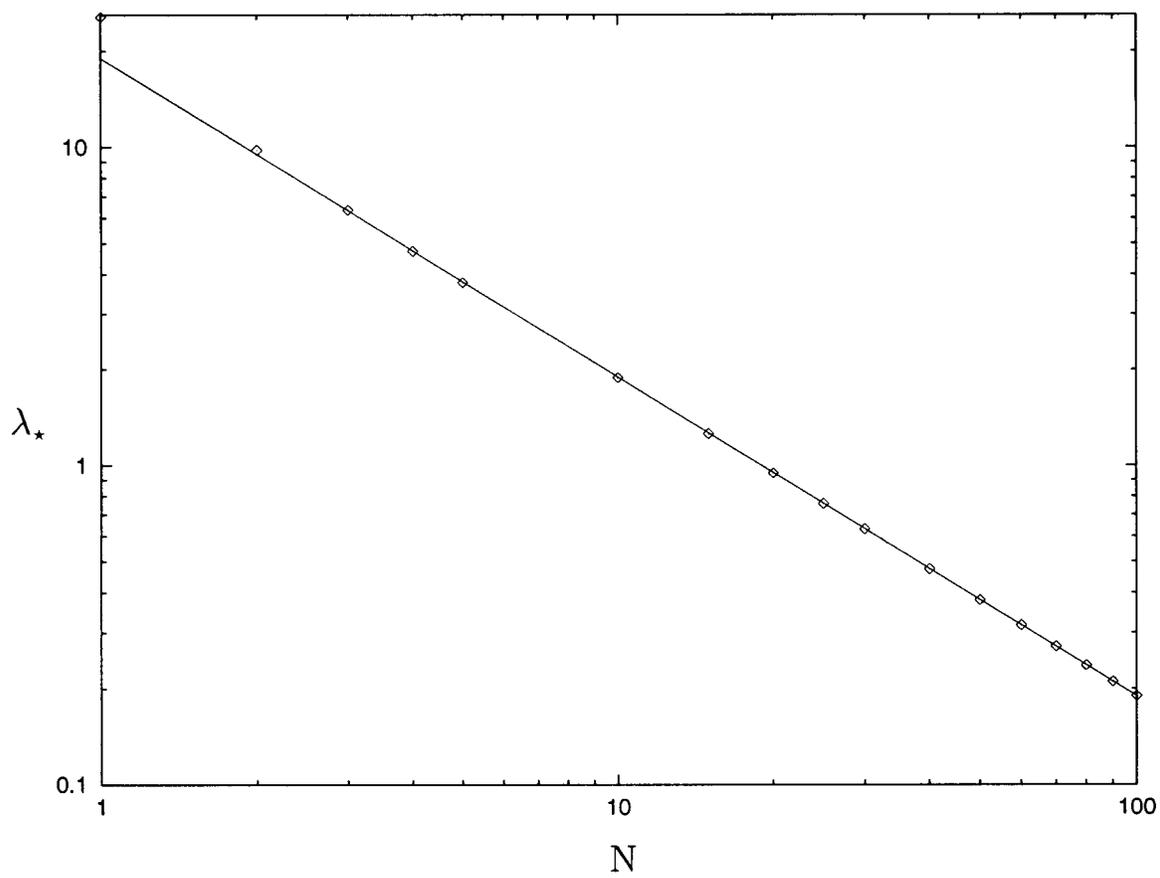
\item[{ Fig.~4:}]  Fixed point value $\lambda_\star$ in dependence on
$N$ for a $\varphi^4$-approximation.
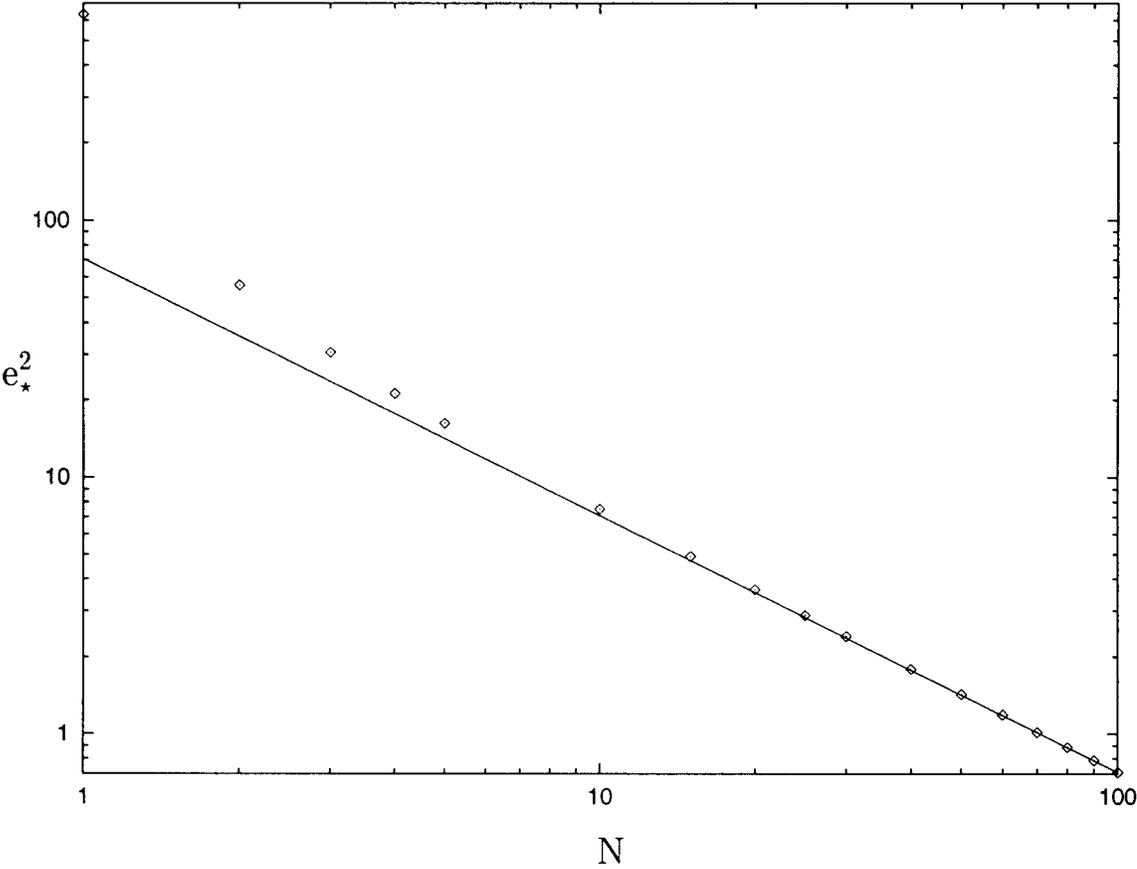
\item[{ Fig.~5:}] Fixed point value $e^2_\star$ in dependence on
$N$ for a $\varphi^4$-approximation.
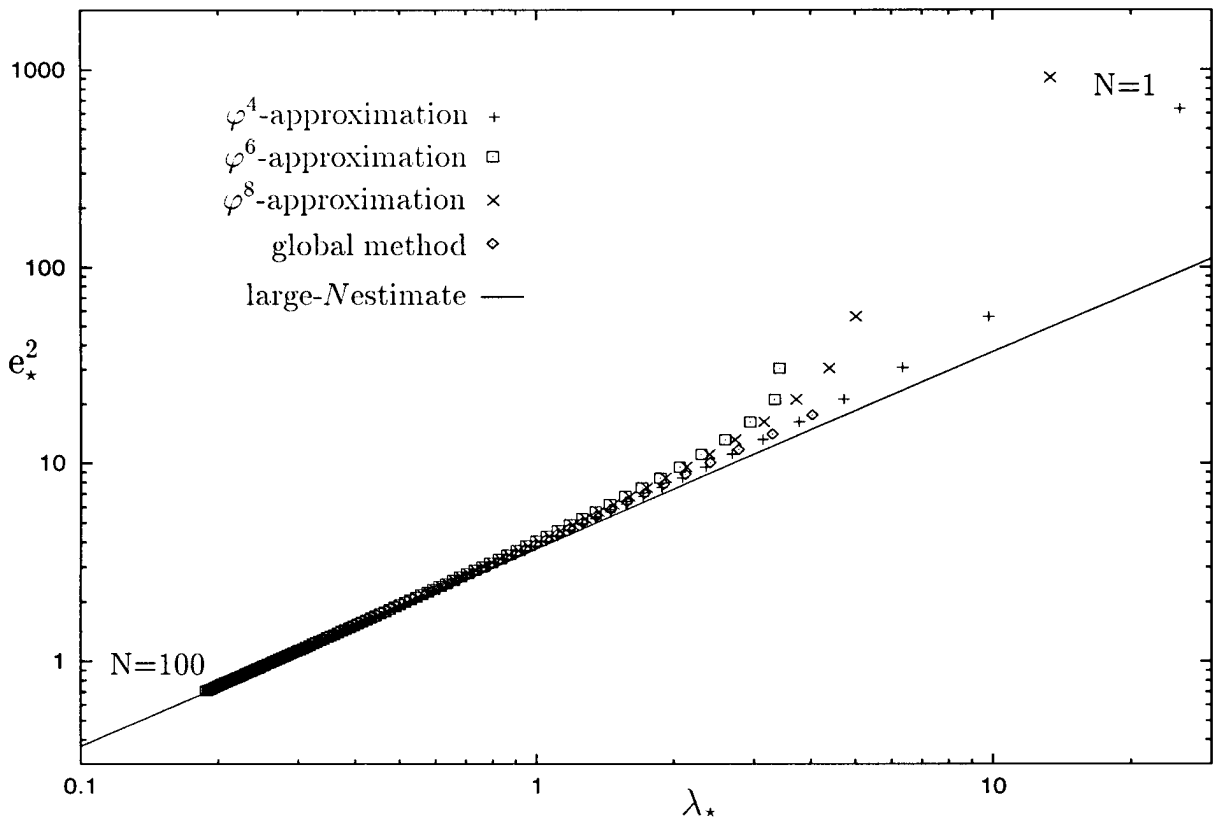
\item[{ Fig.~6:}] Values for $e^2$ and \l
at the second-order fixed point in dependence on $N$
for different approximations.
\item[{ Fig.~7:}] Values for $e^2$ and \l obtained with
the $\varphi^4$ truncation and
the global method for the tricritical fixed point
in dependence on $N$. The solid line
represents the large-$N$ estimate given by \gl{e2latri}.
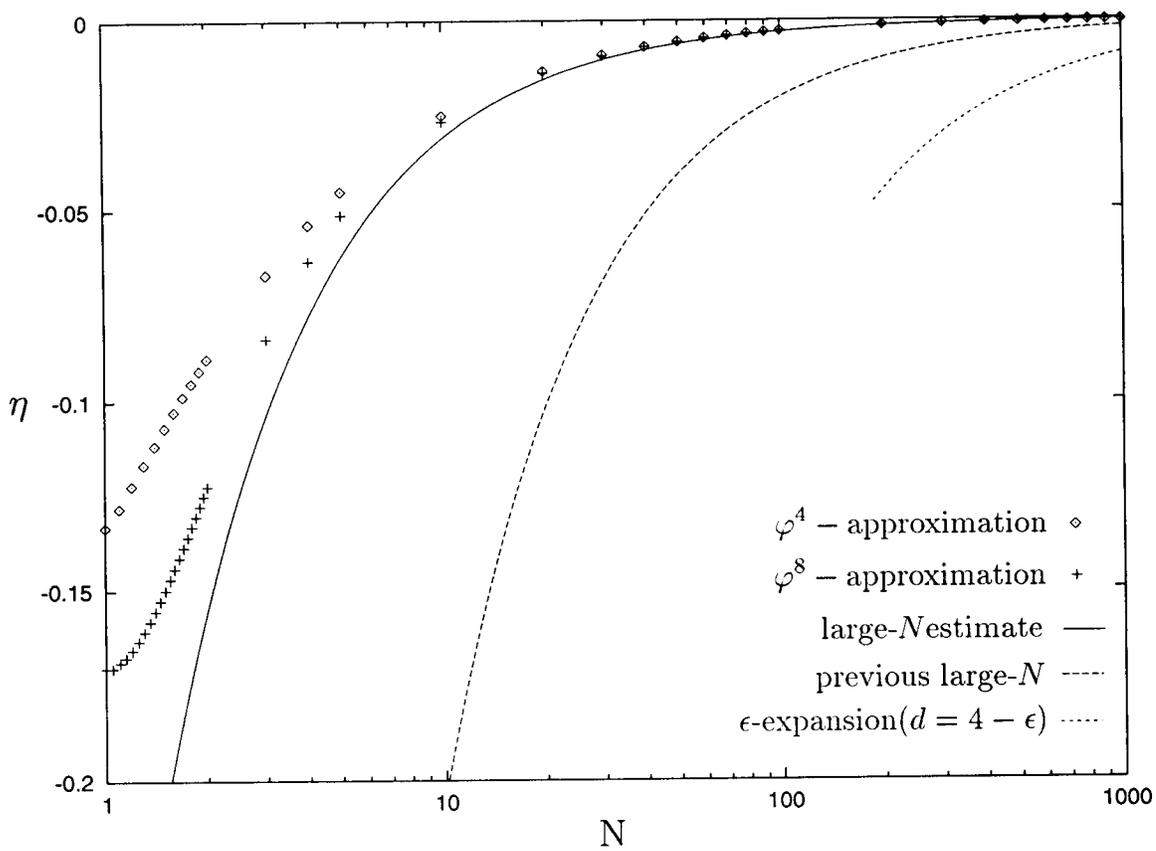
\item[{ Fig.~8:}] The critical index $\eta$  as a function of $N$
for two different approximations and
in comparison with previous estimates.
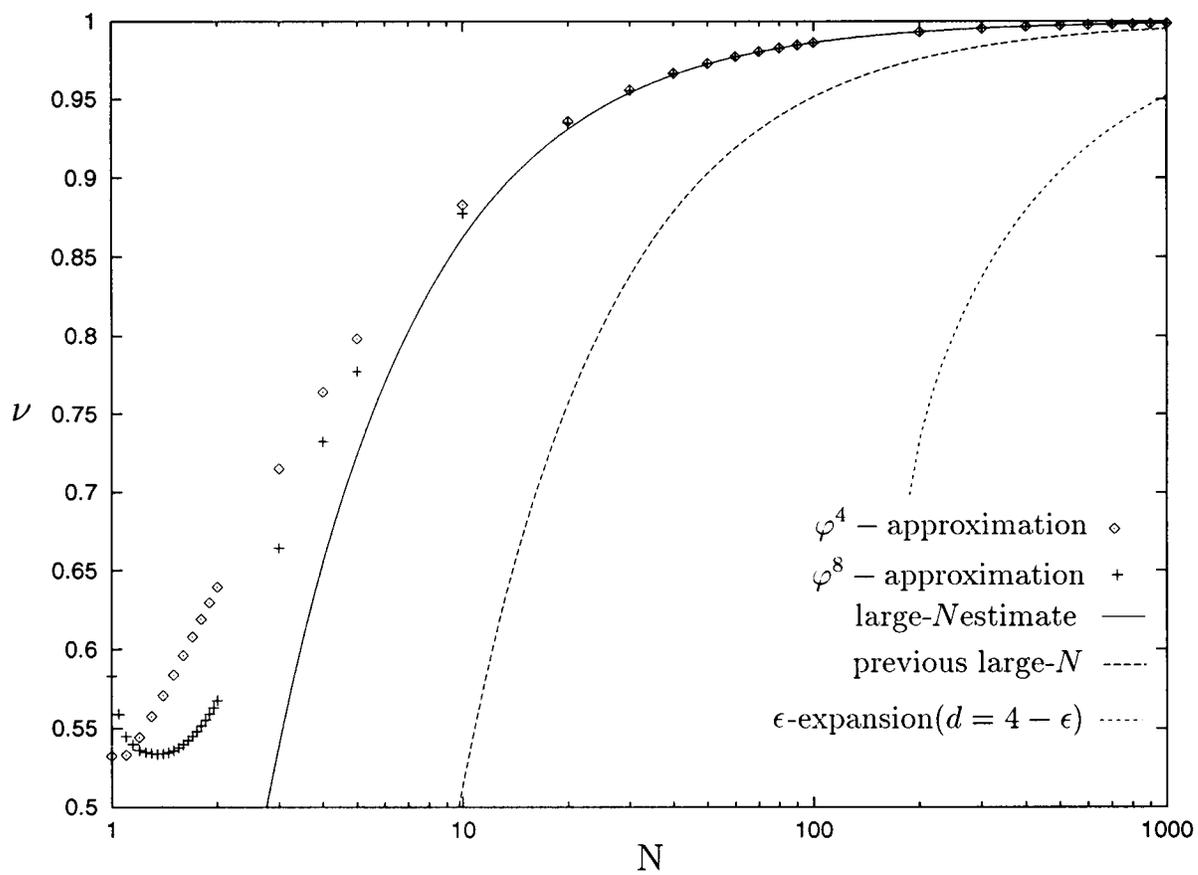
\item[{ Fig.~9:}] The critical index $\nu$ as a function of $N$
for two different approximations and
in comparison with previous estimates.
\end{enumerate}

\begin{thebibliography}{99}

\bibitem{physicaA}
M.\,B.\,Salomon et al, Physica {\bf A200} (1993), 365.

\bibitem{HaLu}
B.\,I.\,Halperin and T.\,C.\,Lubensky, Solid State
Commun. {\bf 14} (1974), 997;\\
P.\,G.\,De Gennes, Solid State Commun. {\bf 14} (1974), 997.

\bibitem{Joetal}
D.\,L.\,Johnson et al, Phys.\,\,Rev. {\bf B18}
(1978), 4902;\\
C.\,W.\,Garland, G.\,B.\,Kasting and K.\,J.\,Lushington,
Phys.\,\,Rev.\,\,Lett {\bf 43} (1979), 1420;\\
C.\,W.\,Garland et al, Phys.\,\,Rev. {\bf A27} (1983), 3234;\\
J.\,Thoen, H.\,Marynissen and W.\,Van Daal, Phys.\,\,Rev.\,\,Lett
{\bf 52} (1984), 204;\\
B.\,M.\,Ocko, R.\,J.\,Birgenean, J.\,D.\,Litster and
M.\,E.\,Neubert,
Phys.\,Rev.\,Lett.
{\bf 52} (1984), 208.

\bibitem{DimRed}
S.\,Weinberg, Phys.\,\,Lett. {\bf B91} (1980), 51; \\
T.\,Appelquist and R.\,Pisarski,
Phys.\,\,Rev. {\bf D23} (1982), 2305; \\
S.\,Nadkarni, Phys.\,\,Rev. {\bf D27} (1983), 917; \\
N.\,P.\,Landsman, Nucl.\,\,Phys. {\bf B322} (1989), 498.

\bibitem{suco}
B.\,I.\,Halperin,
T.\,C.\,Lubensky and S.\,Ma,
Phys. Rev. Lett. {\bf 32}
(1974), 292.

\bibitem{DAS}
C. Dasgupta and B. I. Halperin, Phys. Rev. Lett.
{\bf 47} (1981), 1556; \\
H.\,Kleinert, Lett.\,\,Nuov.\,\,Cim. {\bf 25} (1982), 405;\\
J.\,March-Russell, Phys.\,\,Lett. {\bf B296} (1992), 364.

\bibitem{resum}
J.\,R.\,Espinosa, M.\,Quiros and F.\,Zwirner,
Phys. Lett. {\bf B314} (1993), 206; \\
W.\,Buchm\"{u}ller, Z.\,Fodor, T.\,Helbig
and D.\,Walliser,
Ann. Phys. {\bf 234} (1994), 260; \\
D.\,B\"{o}deker, W.\,Buchm\"{u}ller, Z.\,Fodor
and T.\,Helbig,
Nucl. Phys. {\bf B423} (1994), 171.

\bibitem{ch1}
C.\,Wetterich, Nucl.\,\,Phys. {\bf B352}
(1991), 529; \\
C.\,Wetterich, Z.\,\,Phys. {\bf C57}
(1993), 451; {\bf C60} (1993), 461.

\bibitem{cwe}
C. Wetterich, Phys. Lett. {\bf B301} (1993), 90.

\bibitem{ch2}
N.\,Tetradis and C.\,Wetterich, Nucl.\,\,Phys.
{\bf B398} (1993), 659.

\bibitem{expo}
N.\,Tetradis and C.\,Wetterich,
Nucl.\,\,Phys. {\bf B422} (1994), 541.

\bibitem{ch3}
M.\,Reuter and C.\,Wetterich,  Nucl.\,\,Phys. {\bf B408} (1993), 91;
{\bf B417} (1994), 181;
{\bf B427} (1994), 291.

\bibitem{daniel}
D.\,Litim, C.\,Wetterich and N.\,Tetradis,
HD-THEP-94-23 and OUTP-94-12 preprint.

\bibitem{bastian}
B.\,Bergerhoff and C.\,Wetterich,
HD-THEP-94-31 preprint.

\bibitem{ber}
B.\,Bergerhoff et. al.,
``Phase diagram of superconductors'',
HD-THEP-95-5 preprint.

\bibitem{Amit}
See e.g. D.\,J.\,Amit, {\em Field Theory,
the Renormalization Group, and
Critical Phenomena}, World Scientific,
1984.

\bibitem{hikami}
S.\,Hikami, Prog.\,\,Theor.\,\,Phys. {\bf 62}, No.~1 (1979), 226.

\bibitem{Parisi}
G.\,Parisi, J.\,\,Stat.\,\,Phys. {\bf 23} (1980), 49;
{\em Statistical Field Theory}, Addison Wesley, 1988.

\bibitem{LN}
P.\,Arnold and L.\,Yaffe, Phys.\,\,Rev. {\bf D49}
(1994), 3003.

\bibitem{scalar}
See e.g. J.\,Zinn-Justin, {\em Quantum Field Theory and
Critical Phenomena}, Oxford Science Publications, 1989.

\bibitem{CPN-1}
M.\,L{\"u}scher, Phys.\,\,Lett. {\bf B78} (1978), 465;\\
E.\,Witten, Nucl.\,\,Phys. {\bf B149} (1979), 285.

\bibitem{lattice}
J. Bartholomew, Phys. Rev. {\bf B28}
(1983), 5378; \\
Y. Munehisa, Phys. Lett. {\bf B155} (1985), 159.

\bibitem{nikt}
N. Tetradis, private communication.

\end{thebibliography}
\end{document}